\documentclass[12pt,3p]{elsarticle}
\usepackage{setspace}
\usepackage{amssymb}
\usepackage{rotating}
\usepackage{tabularx} 
\usepackage{float}
\usepackage{booktabs}
\usepackage{url}
\usepackage{lineno}
\usepackage{hyperref}
\hypersetup{colorlinks,linkcolor={[RGB]{0 0 139}},citecolor={[RGB]{0 0 139}},urlcolor={[RGB]{0 0 139}}}
\usepackage{amsmath}
\usepackage{subfigure}
\usepackage{multicol}
\usepackage{wrapfig}
\usepackage{array}
\usepackage[dvipsnames]{xcolor}

\usepackage{soul}  

\newcommand{\volume}{{\ooalign{\hfil$V$\hfil\cr\kern0.08em--\hfil\cr}}}  

\makeatletter
\def\ps@pprintTitle{%
	\let\@oddhead\@empty
	\let\@evenhead\@empty
	\let\@oddfoot\@empty
	\let\@evenfoot\@oddfoot
}
\makeatother
\journal{*************}

\begin{document}
	
	\begin{frontmatter}
		
            \title{Non-homogeneous anisotropic bulk viscosity for \\ acoustic wave attenuation in weakly compressible methods}	
	
		\author[1]{Dheeraj Raghunathan}
		\address[1]{School of Mechanical Sciences\\ Indian Institute of Technology Goa\\ Farmagudi-403401, India}
		
		\author[1]{Y. Sudhakar \corref{cor1}}
		
		\cortext[cor1]{Corresponding author at: School of Mechanical Sciences, Indian Institute of Technology Goa, Farmagudi - 403401, Goa, India. Email: sudhakar@iitgoa.ac.in}
  
\begin{abstract}
    A major limitation of the weakly compressible approaches to simulate incompressible flows is the appearance of artificial acoustic waves that introduce mass conservation errors and lead to spurious oscillations in the force coefficients. In this work, we propose a non-homogeneous anisotropic bulk viscosity term to effectively damp the acoustic waves. By implementing this term in a computational framework based on the recently proposed general pressure equation, we demonstrate that the non-homogeneous and anisotropic nature of the term makes it significantly more effective than the isotropic homogeneous version widely used in the literature. Moreover, it is computationally more efficient than the pressure (or mass) diffusion term, which is an alternative mechanism used to suppress acoustic waves. We simulate a range of benchmark problems to comprehensively investigate the performance of the bulk viscosity on the effective suppression of acoustic waves, mass conservation error, order of convergence of the solver, and computational efficiency. The proposed form of the bulk viscosity enables fairly accurate modelling of the initial transients of unsteady simulations, which is highly challenging for weakly compressible approaches, and to the best of our knowledge, existing approaches can't provide an accurate prediction of such transients.
\end{abstract}
		
		
\end{frontmatter}
	

\section{Introduction}
\label{intro}
Incompressible flows of Newtonian, constant-property fluids are mathematically described by the Navier-Stokes equations given by the following non-dimensional form,
\begin{subequations}
	\begin{align}
    \nabla \cdot \mathbf{u} &= 0,   \label{continuity} \\
    \frac{\partial \mathbf{u}}{\partial t}  + \mathbf{u} \cdot \nabla \mathbf{u} &= - \nabla p + \frac{1}{Re}  \nabla^2 \mathbf{u},  \label{momentum}
    \end{align}
\end{subequations}
where $\mathbf{u}$ and $p$ denote velocity vector and pressure, respectively; $Re$ is the Reynolds number. The major challenge to solve them numerically is the lack of an evolution equation for pressure. The most widely used family of methods~\cite{patankar,Brown1995,perot_1993}, by manipulating the above equations, obtain an equation for pressure or pressure correction~($\phi$) of the following form
\begin{equation}
    \nabla ^2 \phi = \frac{1}{\Delta t} \nabla \cdot \mathbf{u}^*,
\end{equation}
where $\mathbf{u}^*$ represents an intermediate velocity vector. It is computed from the momentum equations but does not satisfy the continuity equation. The numerical solution of such an elliptic equation is highly expensive, requires extensive memory, and offers challenges to achieve scalability on parallel computing architectures~\cite{Borok2007}. 

To avoid solving the expensive Poisson equation, Chorin~\cite{Chorin1967} introduced the concept of pseudo-compressibility and proposed the following pressure evolution equation that replaces the continuity equation
\begin{equation}
   \frac{\partial p}{\partial \tau} + \frac{1}{Ma^2} \nabla \cdot \mathbf{u} = 0.
  \label{pressure_chorin}
\end{equation}
Here, $Ma$ is the Mach number, which is a numerical parameter that quantifies the pseudo-compressibility, and $\tau$ represents the pseudo-time. The above equation, together with the momentum equations, can be used to simulate \textit{steady-state} incompressible flows~\cite{turkel1987,peyret1983}. This technique is called the artificial compressibility method (ACM). The utility of ACM is further extended to unsteady flows by adopting a dual time-stepping~\cite{merkle1987} approach, in which the momentum equations are also appended with a pseudo-time derivative term, as shown below
\begin{equation}
   \frac{\partial \mathbf{u}}{\partial \tau} + \frac{\partial \mathbf{u}}{\partial t} + \mathbf{u} \cdot \nabla \mathbf{u} = - \nabla p + \frac{1}{Re}  \nabla^2 \mathbf{u}.
  \label{dual_time_stepping_momentum}
\end{equation}

Such a modification insists that for every physical time step of the solution process, an extra set of inner iterations needs to be performed until the pseudo-time derivatives diminish to a predefined tolerance. The dual time-stepping approach can be easily ported to a parallel computing framework and is extensively employed to simulate a wide variety of flow problems~\cite{soh1987,rogers1991upwind,Kiris2002,nourgaliev2004,Kim1999,ranjan2020}. However, the presence of the inner iterations makes them computationally expensive. 

The computational cost of the aforementioned approaches to simulate unsteady incompressible flows is dictated by the Poisson equation or pseudo-time iterations. Both these operations are highly time-consuming, and their necessity stems from the unavailability of a separate equation for pressure. In order to circumvent these expensive steps, there has been a growing interest in recent years to derive, from conservation laws, a time-evolution equation for pressure to model incompressible flows. Two notable works that received considerable attention are the entropically damped artificial compressiblity method~(EDAC)~\cite{Clausen2013} and the general pressure equation~(GPE)~\cite{Toutant2017,Toutant2018}. While EDAC uses the entropy balance equation to formulate a thermodynamic constraint equation, GPE is obtained by modifying the energy conservation equation. We underline that the computational frameworks developed based on these equations neither require solving a pressure Poisson equation nor involve dual time schemes. Since they have the potential to be efficiently parallelised on the modern CPU and GPU supercomputing architectures, these methods are in a state of rapid development. Several algorithms, utilizing GPE and EDAC, have been proposed in recent years to handle laminar~\cite{Clausen2013,Toutant2018}, turbulent~\cite{kajzer2018,trojak2022,dupuy2020,Shi2020}, two-phase~\cite{kajzer2020,huang2020arxiv,bodhanwalla2023}, buoyancy-driven~\cite{sharma2023} flows and moving boundary simulations~\cite{bolduc2023}. Another relevant formulation is the kinetically reduced local Navier-Stokes~\cite{Ansumali2005} equation, which does not directly solve for pressure but uses grand potential from which the pressure field is obtained.

In this paper, similar to~\cite{kajzer2018}, \emph{weakly compressible} methods denote techniques that retain some degree of compressibility and hence non-zero divergence of the velocity vector. They include GPE, EDAC, lattice Boltzmann methods and smoothed particle hydrodynamics. ACM, on the other hand, employs dual time-stepping to satisfy the incompressibility condition at each time step. A major limitation of weakly compressible methods is the appearance of artificial acoustic (pressure) waves in the flow field that would have been absent in incompressible flows. Lu and Adams~\cite{lu2023} highlight that there are two mechanisms used in weakly compressible methods to handle acoustic waves: diffusion term in the pressure (or density) equation and bulk viscosity. Toutant~\cite{Toutant2018} conducted a preliminary study on the effect of increasing the amount of pressure diffusion in the GPE to control the acoustic waves. However, we will show that this provides only a limited benefit, which comes at the expense of an additional time-step restriction for stable computations.

The presence of acoustic waves leads to mass conservation errors, extended time to reach a periodic vortex-shedding phase of bluff body flows, highly oscillatory flow-induced forces in truly unsteady flow problems, etc. These limitations need to be addressed in order to make such methods useful in practical applications. We will demonstrate in this paper that the non-homogeneous anisotropic form of bulk viscosity is very effective in addressing all these limitations.

\subsection{Bulk viscosity in weakly compressible methods}
The idea of using bulk viscosity in weakly compressible formulations can be traced back to the work of Ramshaw and Mousseau~\cite{ramshaw1990}. They added the bulk viscosity to improve the convergence rate of the ACM simulations towards the steady state. McHugh and Ramshaw~\cite{mchugh1995} employed bulk viscosity in a dual-time scheme solver with fully implicit discretization. They reported that this term is effective at large time steps but produced no effect on simulations with small time steps. Although the solver is unsteady, they reported results only for steady-state problems. Bulk viscosity has also been used to accelerate the convergence of low Mach number flows to steady state~\cite{ramshaw1991lowMach,mazaheri2003}. A recent study~\cite{yasuda2023} investigated the influence of bulk viscosity on stability and computational efforts in an unsteady simulation using the ACM. For the doubly periodic shear layer problem chosen, they reported that the bulk viscosity suppressed spurious vortices and enabled stable simulations even on coarse grids. Huang~\cite{huang2020arxiv}, in his work on the development of a method to handle two-phase flows using GPE, reported that bulk viscosity is mandatory for stable computation of three-dimensional flows. Recently, AbdulGafoor et al.~\cite{abdulgafoor2024} compared the performance of EDAC and bulk viscosity-enabled classical ACM on static and dynamically distorted grids. They used an isotropic bulk viscosity and observed that the choice of the grid influences the flow field more than the artificial/weakly compressible method used.

The inclusion of bulk viscosity term is not limited to methods that solve Navier-Stokes equations. Its use has been reported in other weakly compressible approaches like Lattice Boltzmann methods~(LBM) and Smoother Particle Hydrodynamics~(SPH) techniques.

Dellar~\cite{dellar2001} reported that the bulk viscosity, when its value is set to be larger than that of the shear viscosity, damped artifacts in unresolved simulations of incompressible flows using LBM. The term acts only in regions of velocity divergence errors, and leaves the other parts affected. Asinari \& Karlin~\cite{asinari2010} proposed an LBM which contains a free parameter to control the second viscosity to enhance the numerical stability in the incompressible limit. They underlined that even though it is not explicitly expressed in the literature, the stability of LBMs in the incompressible can be attributed to the bulk viscosity term.

In the weakly compressible SPH framework, Avalos et al.~\cite{avalos2020} introduced a bulk viscosity term that can be modelled independently of the shear viscosity. The proposed term conserves both linear and angular momentum, which are desirable for simulating flows with free surfaces. In a most recent work, Sun et al.~\cite{sun2023} proposed bulk viscosity as an acoustic damper term in SPH and demonstrated a drastic reduction in pressure oscillations. They rigorously tested the effectiveness of this term on test cases involving violent fluid-fluid and fluid-solid interaction problems.

\subsection{Summary of existing studies and novelty of the present work}
 This section provides a brief summary of existing works on weakly compressible approaches and bulk viscosity. We underline the limitations of these studies and elaborate on the contributions of the present work.
\begin{enumerate}
    \item By analyzing several weakly compressible approaches, Lu and Adams~\cite{lu2023} note that there are two general mechanisms used for stable computations: pressure diffusion (also referred to as mass diffusion) and bulk viscosity. Despite their widespread usage, the relative effectiveness of these two mechanisms remains unclear. Lu and Adams~\cite{lu2023} remarked that both of these are effective. In this paper, we shed light on, for the first time, the effectiveness of these two mechanisms. Our results reveal that the bulk viscosity has a stronger impact in suppressing acoustic waves.
    \item The existing works on introducing the bulk viscosity either addressed its effect on the convergence rate of steady state solvers~\cite{ramshaw1990}, or on the stability~\cite{yasuda2023}. Sun et al.~\cite{sun2023} is the only study that extensively investigated the artificial damping effect but in the SPH framework. In this paper, we present a \emph{comprehensive} study of bulk viscosity on various aspects of an unsteady solver: damping of acoustic waves, order of convergence, and mass conservation error.
    \item The key contribution of the present work is the use of non-homogeneous anisotropic bulk viscosity. These features enable the rapid annihilation of artificial acoustic waves. We will show that this leads to accurate computation of force coefficients on an impulsively started plate, and a drastic reduction in computation cost to model von-Karman vortex shedding behind a bluff body. The above advantages would not have been realized with the isotropic and uniform bulk viscosity used throughout the literature. An earlier technical report~\cite{ramshaw1986} briefly discusses the use of anisotropic but uniform bulk viscosity tensor; however, no numerical tests were presented. To the best of our knowledge, there is no other existing weakly compressible approach that utilises this idea.
\end{enumerate}

The weakly compressible framework we use in this work is built on GPE. This is because Toutant~\cite{Toutant2018} showed that the extra convective term in the EDAC equation offers a negligible influence on the evolution of the flow field.

\subsection{Structure of the paper}
The rest of the paper is arranged as follows. The governing equations and their spatial as well as temporal discretisation details are presented in \S~\ref{numMethods}, with an emphasis on the anisotropic nature and the spatial variation of the bulk viscosity term. A series of numerical experiments conducted to demonstrate the influence of the bulk viscosity term on different aspects of the numerical simulation are summarised in \S~\ref{test_cases}. Conclusions drawn from these studies are described in \S~\ref{conclusion_section}. Some additional discussions are also included in~\ref{appendix_secondTerm} and~\ref{appendix_PrStudy} to supplement our findings.

\section{Governing equations and numerical methods}    
      \label{numMethods}
      
In this section, the governing equations are presented, along with the numerical discretisation details. We use the following two approaches to simulate unsteady incompressible flow problems numerically.

\subsection{GPE without artificial bulk viscosity (denoted as `\normalfont GPE solver')}
GPE is derived from the compressible flow and energy conservation equations in the low Mach number limit~\cite{Toutant2017}, and for an isothermal incompressible flow, GPE is written in non-dimensional form as
\begin{equation}
\frac{\partial p}{\partial t} +\frac{1}{Ma^2} \nabla \cdot \mathbf{u} = \frac{1}{Re Pr} \nabla^{2} p
\label{gpe}
\end{equation}
where, $\mathbf{u} = [\,u \quad  v\,]^{\top}$ is the velocity vector ($u$ and $v$ being its components in a 2D Cartesian coordinate system) and $p$ denotes pressure. The parameters $Ma$ and $Pr$ are the Mach number and Prandtl number, respectively;  $Re$ represents the Reynolds number of the flow. Here, $Ma$ and $Pr$ are the numerical parameters, and $Re$ is the only physical parameter.

Together with GPE, the momentum equations (equation~\eqref{momentum}) govern the evolution of the flow field in the incompressible limit. This approach has been used to simulate various laminar and turbulent incompressible flow problems~\cite{Toutant2018,Shi2020,dupuy2020,pan2022,bodhanwalla2023}.

\subsection{GPE with artificial bulk viscosity (denoted as `\normalfont GPE+BV solver')}
In this approach, we solve the same GPE equation (equation~\eqref{gpe}) to determine pressure. In contrast to the GPE solver, the momentum equation is appended with an artificial bulk viscosity tensor~($\boldsymbol{\mathcal{B}}$). The modified momentum equation is presented below
\begin{equation}
\frac{\partial \mathbf{u}}{\partial t}  + \mathbf{u} \cdot \nabla \mathbf{u} = - \nabla p + \frac{1}{Re}  \nabla^2 \mathbf{u} +
    \nabla \cdot \left( \boldsymbol{\mathcal{B}} \nabla \cdot \mathbf{u} \right)     
\label{momentum_bulkVisc}
\end{equation}
where the artificial bulk viscosity term can be expanded as $\nabla \cdot \left( \boldsymbol{\mathcal{B}} \nabla \cdot \mathbf{u} \right) =  \boldsymbol{\mathcal{B}} \nabla \left( \nabla \cdot \mathbf{u} \right) +    (\nabla \cdot \boldsymbol{\mathcal{B}}) \left(  \nabla \cdot \mathbf{u} \right)$. The second term $(\nabla \cdot \boldsymbol{\mathcal{B}}) \left(  \nabla \cdot \mathbf{u} \right)$ arises due to the non-homogeneous nature of the bulk viscosity as proposed in this work. Note that the above equation is in non-dimensional form;  The dimensionless bulk viscosity is defined as $\boldsymbol{\mathcal{B}}=(\boldsymbol{\mathcal{B}}^*/\rho UL)$, where $\boldsymbol{\mathcal{B}}^*$ is the respective dimensional quantity. In the present work, we propose the following form of the anisotropic bulk viscosity tensor.
\begin{equation}
\boldsymbol{\mathcal{B}} = 
    \begin{bmatrix}
        \mathcal{B}^{\mathcal{X}} & 0 \\
        0 & \mathcal{B}^{\mathcal{Y}}
    \end{bmatrix}
    \label{anisotropic_bulkVisc}
\end{equation}
where $\mathcal{B}^{\mathcal{X}}$ and $\mathcal{B}^{\mathcal{Y}}$ are the components of the tensor in their usual notation. We underline that these components are grid-size dependent and, hence, non-homogeneous in the computational domain. We propose the following functional form of these components
\begin{align}
    \mathcal{B}^{\mathcal{X}}&=\lambda\Delta x, \\
    \mathcal{B}^{\mathcal{Y}}&=\lambda\Delta y, 
\end{align}
where $\lambda>0$ is a constant, $\Delta x$ and $\Delta y$ represent the grid size in $x$ and $y$ directions respectively. In a non-uniform Cartesian mesh, the components of bulk viscosity, for the chosen form, are functions of the respective coordinates only, i.e., $\mathcal{B}^{\mathcal{X}}=f(x)$ and $\mathcal{B}^{\mathcal{Y}}=g(y)$. In a uniform mesh with $\Delta x =\Delta y$, the tensor will be isotropic and uniform, represented by $\boldsymbol{\mathcal{B}}=\mathcal{B} \bf{I}$ where $\mathcal{B}$ is a scalar and $\bf{I}$ is the identity tensor. We use the boldface symbol $\boldsymbol{\mathcal{B}}$ to denote bulk viscosity tensor in a non-uniform mesh and $\mathcal{B}$ to denote its scalar value on a uniform mesh. We emphasize while the bulk viscosity is a physical parameter for compressible flows, it is a free parameter in our work, and hence, its values can be set independent of the shear viscosity~\cite{avalos2020}.

The choice of the functional form for $\boldsymbol{\mathcal{B}}$ is driven by the following argument. To completely eliminate artificial waves, the time scale associated with the bulk viscosity should be much smaller than the acoustic time scale. For a 1D flow, in non-dimensional settings, this implies (refer to equations~\eqref{time_step_eqn_Ma} and ~\eqref{time_step_eqn_B})
\begin{equation}
    \frac{\Delta x^2}{\mathcal{B}}\ll \frac{Ma\Delta x}{U}.
\end{equation}
However, ensuring the above condition requires a reduction in the time step of the simulation, which leads to a significant increase in computational cost. So, our objective is to introduce maximum diffusion of acoustic waves without introducing additional restrictions on the time step. This leads to
\begin{equation}
    \gamma\frac{\Delta x^2}{\mathcal{B}}=\frac{Ma\Delta x}{U},
\end{equation}
where $\gamma$ is a positive constant, such that $0\le\gamma\le 1$, depending on the problem. From the above, we get
\begin{equation}
    \mathcal{B}=\frac{\gamma U\Delta x}{Ma},
    \label{bulkViscFunctionEqn}
\end{equation}
and by defining $\lambda=\gamma U/Ma$, we obtain the expression defined earlier, $\mathcal{B}=\lambda\Delta x$.

\begin{figure}[!ht]
   \centering
   \includegraphics[trim = 0mm 0mm 0mm 0mm, clip, width=15cm]{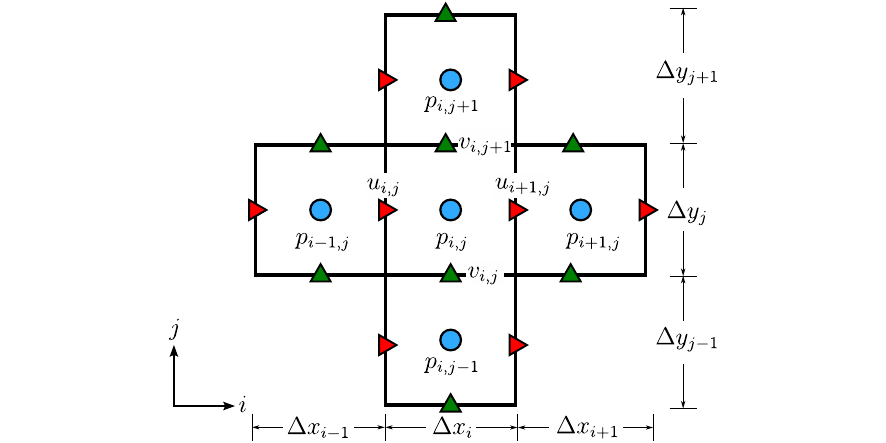}
   \caption{The computational stencil, indicating the arrangement of variables on a staggered grid.}
   \label{staggered}
\end{figure}
 
\subsection{Spatial Discretization}
The governing equations are recast into integral form, and a finite volume approach is adopted to discretize the governing equations. We use the Cartesian coordinate system in which the axes are denoted by $x$ and $y$. Figure \ref{staggered} shows the computational stencil used for discretisation. In order to avoid odd-even decoupling, the variables are staggered. Pressure is stored at the cell centre, and the velocity components are on the cell face centres. All spatial derivatives appearing in the governing equations are approximated using the central difference scheme similar to~\cite{Toutant2018,dupuy2020}. The convective terms are treated in their conservative form. In the following, we present only the discretization details of the bulk viscosity term.
\begin{enumerate}
    \item $x$-component of $\boldsymbol{\mathcal{B}} \nabla \left( \nabla \cdot \mathbf{u} \right)$
\begin{flalign*}
\int_\volume \mathcal{B}^\mathcal{X}  \frac{\partial }{\partial x} (\nabla \cdot \mathbf{u})  d\volume 
\approx   & \frac{\mathcal{B}^\mathcal{X}_{i}}{(0.5\Delta x_i + 0.5\Delta x_{i-1})} \\\biggl[  \left( \frac{u_{i+1,j} - u_{i,j}}{\Delta x_i} +  \frac{v_{i,j+1} - v_{i,j}}{\Delta y_j} \right)  - 
 &   \left( \frac{u_{i,j} - u_{i-1,j}}{\Delta x_{i-1}} +  \frac{v_{i-1,j+1} - v_{i-1,j}}{\Delta y_j} \right) \biggr] \Delta \volume^\mathcal{X}_{i,j}
\end{flalign*}
where,  $\mathcal{B}^\mathcal{X}_{i}$ $= \lambda (0.5\Delta x_i + 0.5\Delta x_{i-1})$ and $\Delta \volume^\mathcal{X}_{i,j} = (0.5\Delta x_i + 0.5\Delta x_{i-1})\Delta y_j $. \\

    \item $y$-component of $\boldsymbol{\mathcal{B}} \nabla \left( \nabla \cdot \mathbf{u} \right)$
\begin{flalign*}
\int_\volume \mathcal{B}^\mathcal{Y}  \frac{\partial }{\partial y} (\nabla \cdot \mathbf{u})  d\volume 
\approx & \frac{\mathcal{B}^\mathcal{Y}_{j}} {(0.5\Delta y_j + 0.5\Delta y_{j-1})} \\ \biggl[  \left( \frac{u_{i+1,j} - u_{i,j}}{\Delta x_i} +  \frac{v_{i,j+1} - v_{i,j}}{\Delta y_j} \right)  - 
 &   \left( \frac{u_{i+1,j-1} - u_{i,j-1}}{\Delta x_{i}} +  \frac{v_{i,j} - v_{i,j-1}}{\Delta y_{j-1}} \right) \biggr] \Delta \volume^\mathcal{Y}_{i,j}
\end{flalign*}
where, $\mathcal{B}^\mathcal{Y}_{j}$ $= \lambda (0.5\Delta y_j + 0.5\Delta y_{j-1})$ and $\Delta \volume^\mathcal{Y}_{i,j} = \Delta x_i (0.5\Delta y_j + 0.5\Delta y_{j-1})$. \\

    \item $x$-component of $(\nabla \cdot \boldsymbol{\mathcal{B}}) \left(  \nabla \cdot \mathbf{u} \right)$
\begin{align*}
\int_\volume   \frac{\partial \mathcal{B}^\mathcal{X} }{\partial x} (\nabla \cdot \mathbf{u})  d\volume 
\approx  \biggl[ & \left( \frac{\lambda \Delta x_i - \lambda \Delta x_{i-1}}{0.5\Delta x_i + 0.5\Delta x_{i-1}}  \right)  \left( \nabla \cdot \mathbf{u}\right)^{\mathcal{X}}_{i,j}  \biggr] \Delta \volume^\mathcal{X}_{i,j}
\end{align*}
where, 
\begin{align*}
\left( \nabla \cdot \mathbf{u}\right)^{\mathcal{X}}_{i,j}  =  &\left( \frac{u_{i+1,j} - u_{i-1,j}}{\Delta x_{i} + \Delta x_{i-1}} \right) + \\ 
 \frac{1}{\Delta y_j} & \left( \frac{v_{i,j+1} \Delta x_{i-1} + v_{i-1,j+1}\Delta x_{i}}{\Delta x_i + \Delta x_{i-1}} - \frac{v_{i,j} \Delta x_{i-1} + v_{i-1,j}\Delta x_{i}}{\Delta x_i + \Delta x_{i-1}}  \right)
\end{align*}
\\
    \item $y$-component of $(\nabla \cdot \boldsymbol{\mathcal{B}}) \left(  \nabla \cdot \mathbf{u} \right)$
\begin{align*}
\int_\volume   \frac{\partial \mathcal{B}^\mathcal{Y} }{\partial y} (\nabla \cdot \mathbf{u})  d\volume 
\approx  \biggl[ & \left( \frac{\lambda \Delta y_j - \lambda \Delta y_{j-1}}{0.5\Delta y_j + 0.5\Delta y_{j-1}}  \right)  \left( \nabla \cdot \mathbf{u}\right)^{\mathcal{Y}}_{i,j}  \biggr] \Delta \volume^\mathcal{Y}_{i,j}
\end{align*}
where, \begin{align*}
\left( \nabla \cdot \mathbf{u}\right)^{\mathcal{Y}}_{i,j}  = \frac{1}{\Delta x_i} & \left( \frac{u_{i+1,j-1} \Delta y_{j} + u_{i+1,j}\Delta y_{j-1}}{\Delta y_j + \Delta y_{j-1}} - \frac{u_{i,j-1} \Delta y_{j} + u_{i,j}\Delta y_{j-1}}{\Delta y_j + \Delta y_{j-1}}  \right) + \\
    & \left( \frac{v_{i,j+1} - v_{i,j-1}}{\Delta y_{j} + \Delta y_{j-1}}  \right)
\end{align*}

\end{enumerate}

Here, $i$ and $j$ indicate the indices in $x$ and $y$ directions, respectively, whereas $\Delta x_i$ and $\Delta y_j$ represent the grid size of a particular control volume as shown in figure~\ref{staggered}.  

\subsection{Temporal Discretization}
We apply the three-stage Strong Stability Preserving Runge-Kutta~\cite{Parameswaran2019} method for the time integration. This procedure is briefly explained below for an ordinary differential equation with the dependent variable $\Phi$. The momentum equation and the GPE can be recast into the following form after the spatial discretization
\begin{equation}
  \frac{d \Phi}{d t}  = \mathit{\mathbb{L}}(\Phi),
  \label{dualacm_momentum_modified}
\end{equation}
where $\mathit{\mathbb{L}}$ is the spatial discretisation operator. We follow the steps below to update $\Phi$ from time level $n$ to $n+1$.
\begin{linenomath}
    \begin{align}
        \begin{split}
            \Phi^{(1)} =& \:\:\: \Phi^{(n)} + \frac{\Delta t}{\volume} \mathit{\mathbb{L}}(\Phi^{(n)}) \\
            \Phi^{(2)} =& \:\:\: \frac{3}{4}\Phi^{(n)} + \frac{1}{4}\Phi^{(1)} + \frac{1}{4} \frac{\Delta t}{\volume} \mathit{\mathbb{L}}(\Phi^{(1)}) \\
            \Phi^{(n+1)} =& \:\:\: \frac{1}{3}\Phi^{(n)} + \frac{2}{3}\Phi^{(2)} + \frac{2}{3} \frac{\Delta t}{\volume} \mathit{\mathbb{L}}(\Phi^{(2)})  
        \end{split}
    \end{align}
\end{linenomath}
The time-step $\Delta t$ for the simulations is determined by considering the diffusive stability requirements and the Courant– Friedrichs–Lewy~(CFL) condition arising from the convective velocity and the artificial acoustic wave speed~\cite{Shi2020,yang2021}. Moreover, the inclusion of the bulk viscosity tensor in the GPE+BV solver adds another restriction on the time step.  Theoretically, for two-dimensional simulations, a minimum over the following bounds determines the time step, as given below.
\begin{equation}
    \Delta t = f_s\cdot\textrm{min} (\Delta t_{acs}, \Delta t_{ad}, \Delta t_{diff}, \Delta t_{visc}, \Delta t_{bv}),
\end{equation}
where $0<f_s\le 1$ is the safety factor. The criteria appearing in the above equation are listed here.
\begin{enumerate}
	\item CFL criterion based on the acoustic wave speed
	\begin{equation}
		\Delta t_{acs}  = \min\limits_{i,j} \left [  \frac{U/Ma}{\Delta x_i} + \frac{U/Ma}{\Delta y_j}   \right ]^{-1}
            \label{time_step_eqn_Ma}
	\end{equation}
	\item CFL criterion based on the convective velocity
	\begin{equation}
	\Delta t_{ad} = \min\limits_{i,j} \left [  \frac{|u_{i,j}|}{\Delta x_i} + \frac{|v_{i,j}|}{\Delta y_j}   \right ]^{-1}
	\end{equation}
 	\item Stability criterion based on the viscous term in the momentum equation
	\begin{equation}
		\Delta t_{visc} = 
		\min\limits_{i,j} \left [ \frac{1}{2}\frac{\Delta_{i,j}^2 }{(1/Re)} \right ]
	\end{equation}
	\item Stability criterion based on the diffusion term in the GPE
	\begin{equation}
		\Delta t_{diff} = 
		\min\limits_{i,j} \left [ \frac{1}{2} \frac{\Delta_{i,j}^2  }{(1/Re Pr)}\right ]
		\label{time_step_eqn_Pr}
	\end{equation}
	\item Time-step restriction due to the bulk viscosity term
	\begin{equation}
		\Delta t_{bv} = 
		\min\limits_{i,j} \left [ \frac{1}{2}\frac{\Delta_{i,j}^2 }{\mathcal{B}} \right ]
		\label{time_step_eqn_B}
	\end{equation}
\end{enumerate}
In the above expressions, $\min\limits_{i,j}$ represents the minimum of all values computed for each cell on the mesh. Note that all the parameters in the above equations are in non-dimensional form. $U$ denotes the characteristic velocity of the flow and $\Delta$ represents the grid-size factor for 2D grids given by, 
\begin{equation*}
    \Delta^2_{i,j} = \frac{\Delta x_i^2 \Delta y_j^2 }{\Delta x_i^2 + \Delta y_j^2}.
\end{equation*}

From the above equations, it is evident that the time-step depends on the flow parameter $Re$ and the three numerical parameters such as $Ma$, $Pr$ and $\mathcal{B}$. Unlike $Re$, which is problem-dependent, the numerical parameters can be adjusted. For weakly compressible approaches, $Ma\ll 1$ so as to closely represent incompressible flows. Hence, $\Delta t_{acs}$ in most simulations dictates the time step of the simulation. A lower value of $Pr$ helps in damping the acoustic waves~\cite{Toutant2018} but at the cost of a lower time-step requirement. Similarly, a large value of $\mathcal{B}$ implies better damping, but requires a reduction in $\Delta t$. Hence, there is a trade-off between the computational cost and the rate of acoustic damping. In this work, we chose a large permissible value of $\mathcal{B}$ for the given allowable $\Delta t_{acs}$. This ensures the maximum benefit of including the bulk viscosity without increasing the computational cost.

\section{Numerical experiments}
\label{test_cases}
The present paper aims to comprehensively investigate the effect of artificial bulk viscosity on weakly compressible methods. In this regard, we present a few benchmark examples covering free-shear flows and flows past a solid
body/obstacle. For each example, we perform simulations without~(denoted as GPE solver) and with the artificial bulk viscosity~(denoted as GPE+BV solver).

It is to be noted that the addition of the bulk viscosity modifies the governing equations as can be seen from equation~\eqref{momentum_bulkVisc}, and hence, one can expect that this leads to a noticeable departure from incompressible flows. However, the addition of $\mathcal{B}$ affects only the acoustic component of pressure, leaving the incompressible part unaffected~\cite{sun2023}. It has been shown indirectly that the bulk viscosity does not significantly affect the velocity fields of single-phase~\cite{lu2023,dellar2001,yasuda2023} and two-phase flows~\cite{kajzer2022, huang2020arxiv}. However, the pressure fields differ due to the presence of artificial acoustic waves in the weakly compressible approaches, until these waves are damped. It is known that these artificial waves dissipate faster with the addition of bulk viscosity. One of the main focuses of this section is to provide evidence for enhanced attenuation of such waves with the proposed anisotropic non-homogeneous version of the bulk viscosity.

Each test case reported in this section investigates a different influence of the inclusion of bulk viscosity. We start with the simple lid-driven cavity problem in order to provide evidence for the damping of acoustic waves by the bulk viscosity. Then, we discuss how the present approach yields reduced mass conservation error using the doubly periodic shear layer (DPSL). Further, we present the simulation data from the Taylor-Green vortex (TGV) problem to showcase how the inclusion of this additional term helps to obtain a uniform order of convergence. Next, we demonstrate for the flow over a square cylinder that the proposed form of the bulk viscosity term helps us to quickly reach the periodic vortex shedding, thereby achieving improved computational efficiency. Finally, for an impulsively started plate, a challenging unsteady problem for weakly compressible approaches, we provide evidence for better force computation using the proposed approach. It is to be noted that the first three problems are simulated on uniform meshes, on which the bulk viscosity is a scalar; the last two test cases are on stretched meshes, and the advantages of using the tensorial form of the bulk viscosity are underlined.

Toutant~\cite{Toutant2018} reported that by setting $Ma=0.02$ and $Pr=1$, GPE produced accurate results for incompressible flows. Unless otherwise stated, we use these values in our simulations. 


\subsection{Lid-driven cavity flow: Evidence of damping by bulk viscosity}
\label{lidCavitySubSection}
\begin{figure}[!ht]
	\centering
	\includegraphics[trim = 0mm 0mm 0mm 0mm, clip, width=15cm]{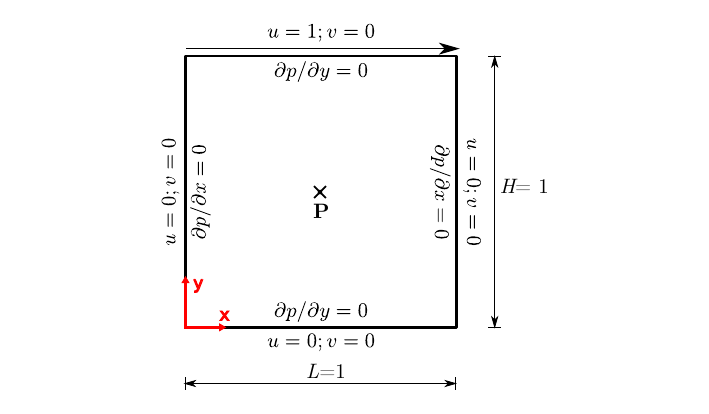}
	\caption{Geometric specifications of lid-driven cavity flow problem. \textbf{P} is the point at which the evolution of pressure with respect to time is studied. It is located at $x=L/2$ and $y=H/2$.}
	\label{cavitySchematic}
\end{figure}

We illustrate the presence of the acoustic pressure waves and their suppression using the bulk viscosity within the well-known lid-driven cavity problem. Although this is a steady-state problem, our objective is to observe the evolution of the flow field until the steady-state is reached. The schematic of the problem and the boundary conditions are shown in figure~\ref{cavitySchematic}. Throughout the domain, both the velocity and pressure fields are initialised to zero. A grid of $64\times 64$ and a time-step of $10^{-4}$ is used. We performed simulations at two different Reynolds numbers, viz. $Re=100$ and $Re=400$. 

\begin{figure}[!ht]
	\centering
	\subfigure[]
	{
		\includegraphics[trim = 0mm 0mm 0mm 0mm, clip, width=7.5cm]{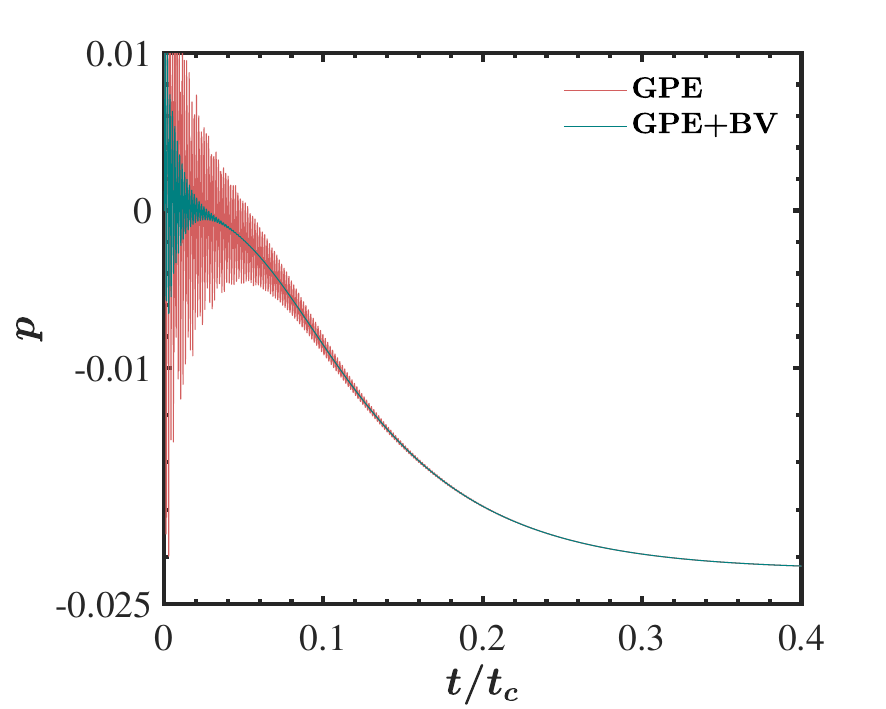}
	}
	\subfigure[]
	{
		\includegraphics[trim = 0mm 0mm 0mm 0mm, clip, width=7.5cm]{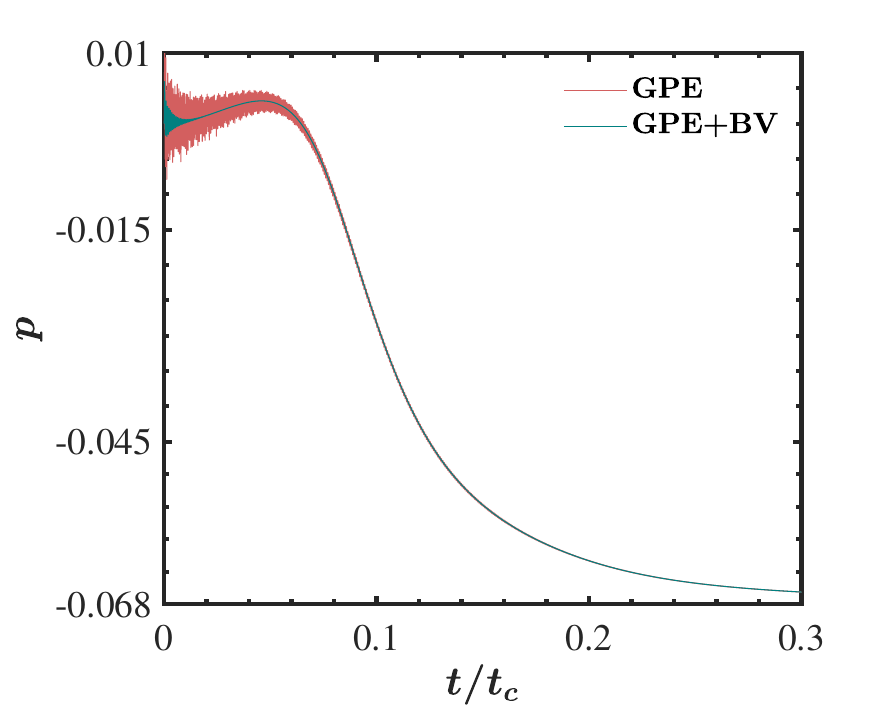}
	}
	\caption{Comparison of pressure evolution at the centre point of the domain, with GPE and GPE+BV solvers for the lid-driven cavity flow problem (a) $Re=100$ (b) $Re=400$.}
	\label{cavity_pressure}
\end{figure} 

Firstly, we simulated the flow without considering the artificial bulk viscosity term. i.e. with the GPE solver. Figure~\ref{cavity_pressure} shows the time variation of the pressure at the cavity centre (denoted by point \textbf{P} in figure~\ref{cavitySchematic}). In these plots, the time is scaled with $t_c$, which is the time required to achieve the steady state. The highly oscillatory nature of the pressure field, seen in the figure, can be attributed to the inherent acoustic waves associated with the weakly compressible family of methods. It is clear that oscillations of high amplitude are visible at the beginning of the simulation, which die out eventually. 

The cavity problem was simulated again with the GPE+BV solver to check the effect of incorporating the artificial bulk viscosity term. The pressure plots, thus obtained, are presented in figure~\ref{cavity_pressure}. We used $\mathcal{B}=50 \Delta x$, with $\Delta x$ being the size of the uniform mesh. Based on the non-dimensional bulk viscosity $\mathcal{B}$ defined earlier, we can interpret the parameter $\mathcal{B} Re$ to provide the ratio of the bulk viscosity to the dynamic viscosity. For this simulation, $\mathcal{B} Re = 78.125$. We observe a rapid damping of pressure oscillations with this minor modification of the solver. When the bulk viscosity term is ignored, the presence of oscillations is visible even at $t/t_c=0.3$ for $Re=100$. In contrast, with the GPE+BV solver, complete damping is attained at around $t/t_c=0.05$. Similarly, for $Re=400$, complete damping is observed at $t/t_c=0.3$ and $t/t_c=0.03$ for GPE and GPE+BV solvers, respectively. A previous study~\cite{ramshaw1990} mentioned such acoustic wave annihilation using bulk viscosity in the lid-driven cavity problem. However, they did not present any pressure evolution plots to visualise the damping.

\begin{figure}[!ht]
	\centering
	\subfigure[]
	{
		\includegraphics[trim = 0mm 0mm 0mm 0mm, clip, width=7.5cm]{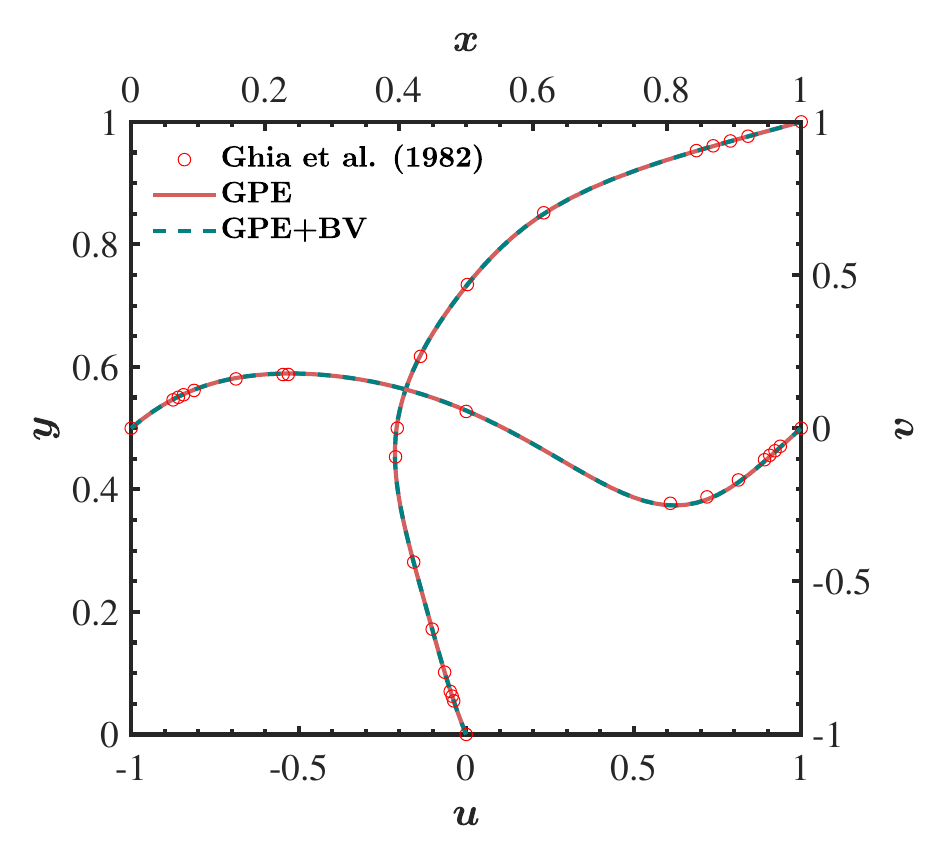}
	}
	\subfigure[]
	{
		\includegraphics[trim = 0mm 0mm 0mm 0mm, clip, width=7.5cm]{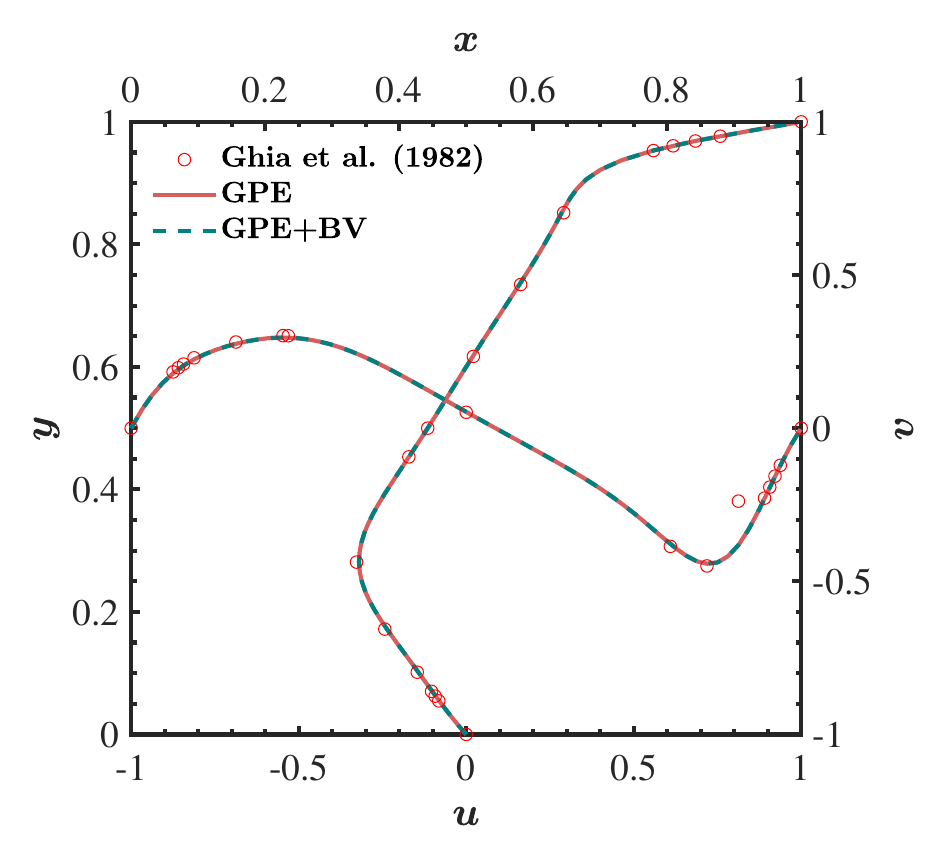}
	}
	\caption{Velocity proﬁles $u$(0.5, $y$) and $v$($x$ , 0.5) for the lid-driven cavity flow problem. (a)~$Re=100$ (b)~$Re=400$.}
	\label{cavity_velocity}
\end{figure}

To ensure accuracy is maintained with the addition of artificial bulk viscosity, we compared velocity profiles from both solvers with those of Ghia et al.~\cite{Ghia1982}. It can be seen in figure~\ref{cavity_velocity} that the plots from GPE and GPE+BV solvers overlap, and they match well with the reference results. This indicates the velocity profiles are intact, and $\mathcal{B}$ only modifies the pressure field.

\begin{figure}[!ht]
	\centering
	\subfigure[]
	{
		\includegraphics[trim = 0mm 0mm 0mm 0mm, clip, width=7.5cm]{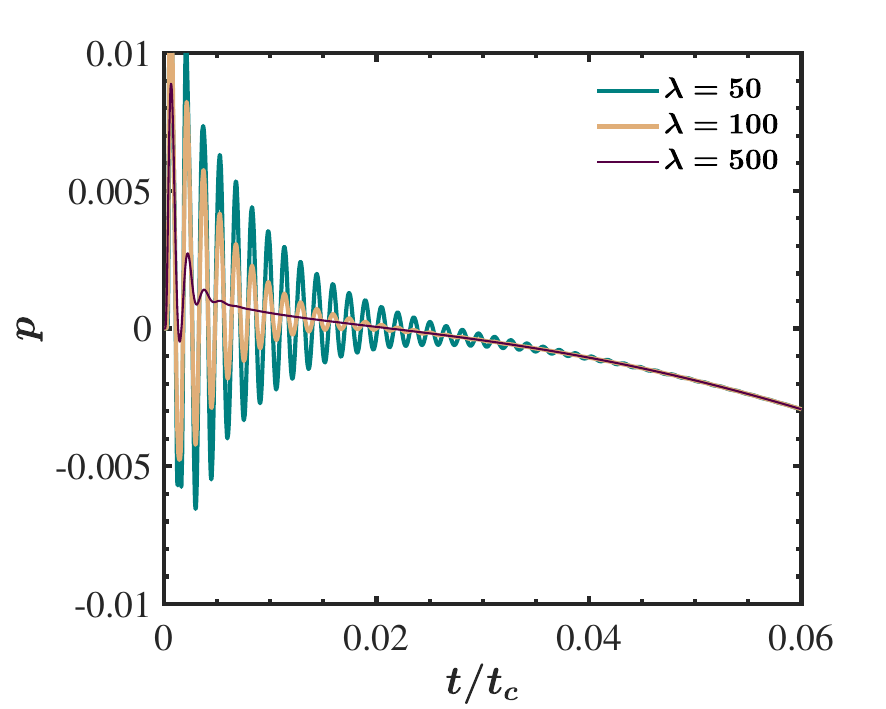}
	}
	\subfigure[]
	{
		\includegraphics[trim = 0mm 0mm 0mm 0mm, clip, width=7.5cm]{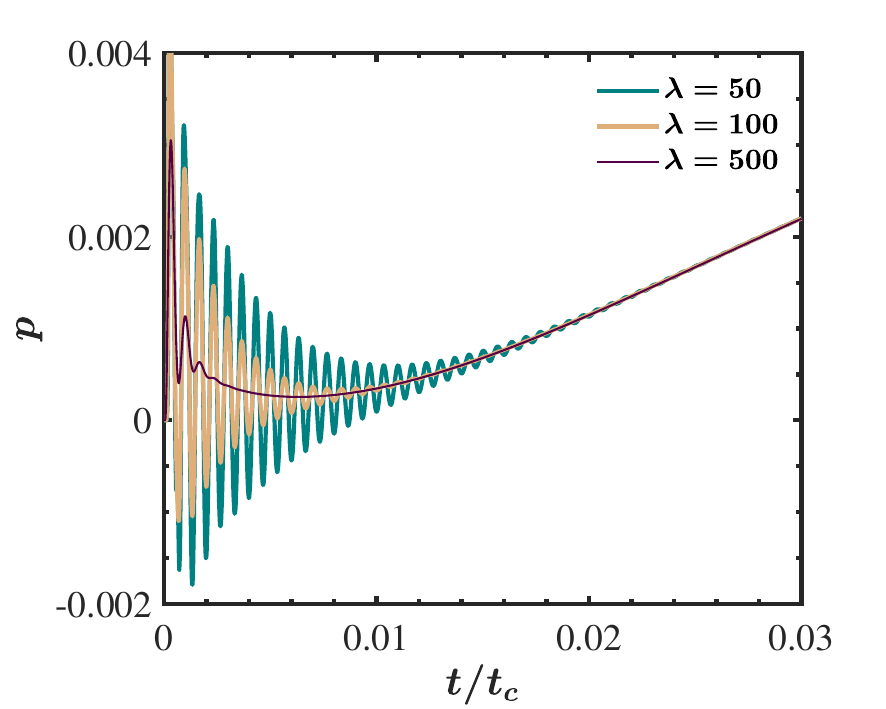}
	}
	\caption{ Effect of artificial bulk viscosity parameter ($\lambda$) for the lid-driven cavity flow problem (a)~$Re=100$ (b)~$Re=400$.}
	\label{cavity_pressure_coeffStudy}
\end{figure} 

It is of interest to investigate the influence of $\mathcal{B}$ on the effectiveness of damping. Since $\Delta x$ is constant throughout the domain, we changed $\lambda$ to vary $\mathcal{B}$. The time-evolution of pressure at the cavity centre for different values of $\lambda$ are presented in figure~\ref{cavity_pressure_coeffStudy}. It is directly evident that increasing the value of $\lambda$ leads to the elimination of oscillations at a faster pace. However, as can be seen from equation~\eqref{time_step_eqn_B}, the allowable $\Delta t$ is inversely proportional to $\mathcal{B}$. For the current problem, setting $\lambda=50$ does not show any stability issues for the chosen time step. However, for  $\lambda=100$ and  $\lambda=500$, we need to reduce the time-step to $10^{-5}$ and $10^{-6}$, respectively. This increases the computational cost but provides a quicker damping rate. Hence, the value of $\mathcal{B}$ should be based on a pay-off between the computational expenses and the rate of damping the acoustic pressure waves. In all other simulations, we choose $\mathcal{B}$ so that the simulation is stable for $\Delta t $ decided by the $Ma$ as given in equation~\eqref{time_step_eqn_Ma}. This is done so as to get the maximum benefit of acoustic wave suppression without introducing additional computational overhead by reducing the time step.

Pressure oscillations in the lid-driven cavity flow have been presented previously using similar weakly compressible flow solvers such as classical ACM~\cite{he2002} and lattice Boltzmann methods~\cite{he2002,nagata2021}. Similar pressure plots presented by~\cite{nagata2021} indicate that the pressure oscillations are visible even at $t/t_c=0.7$ for $Re=100$, whereas for GPE and GPE+BV solvers, the oscillations are eliminated at $t/t_c=0.3$ and 0.05, respectively. We can conclude from this that the pressure diffusion term in GPE provides damping of acoustic waves as envisaged by Toutant~\cite{Toutant2017}, but the present modification enhances the damping significantly.


\begin{figure}[!ht]
    \centering
    \includegraphics[trim = 0mm 0mm 0mm 0mm, clip, width=15cm]{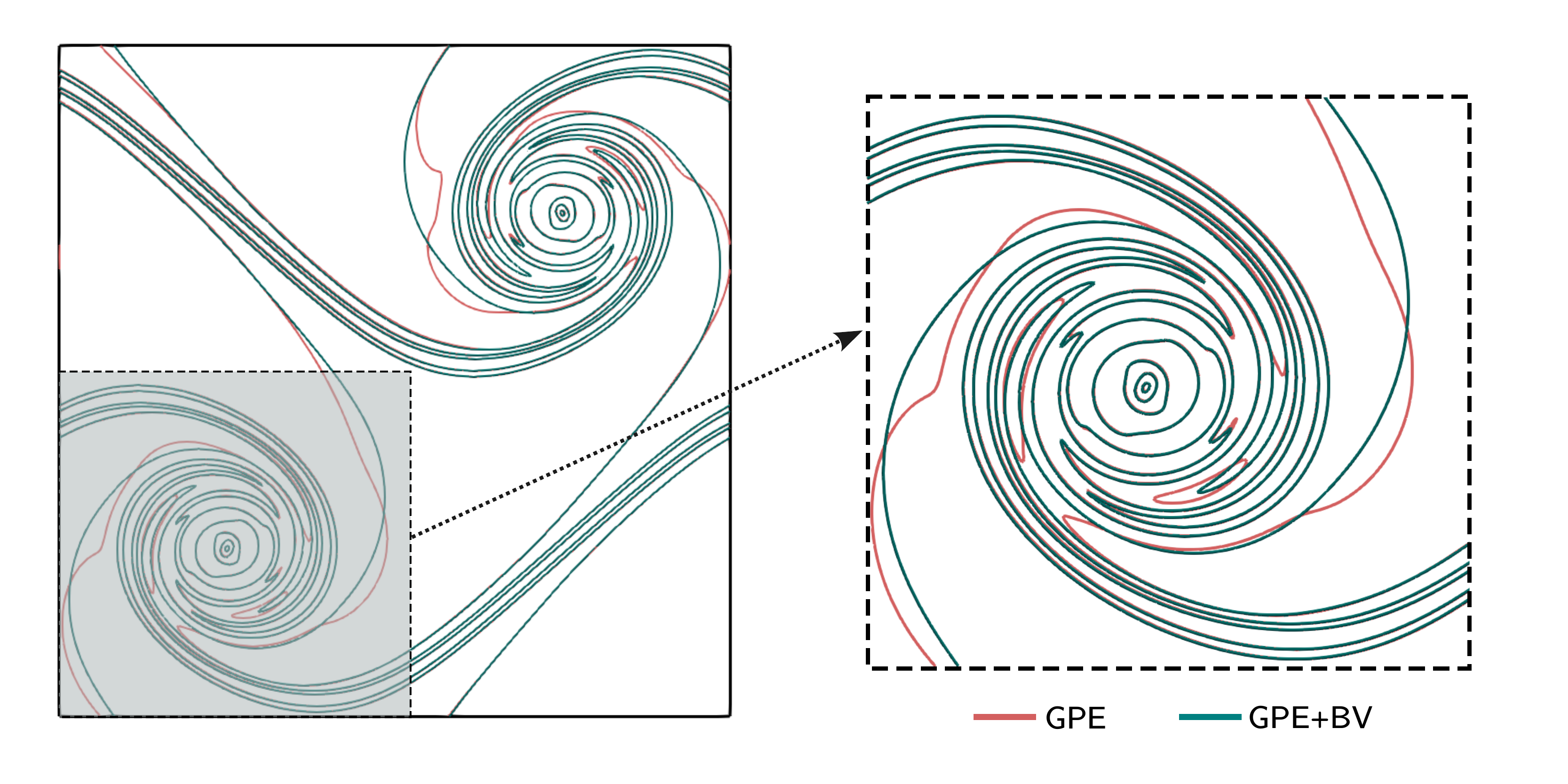}
    \caption{Comparison of vorticity contours from GPE and GPE+BV solvers for the doubly periodic shear layer test case with a grid of $512 \times 512$ at $t=1$. The vorticity contours are plotted for $\omega_z = 0, \pm6, \pm18, \pm30, \pm45, \pm55, \pm57, \pm57.5$.}
    \label{dpsl_vortComparison}
\end{figure}

\subsection{Doubly periodic shear layer: Effect on velocity divergence}
\label{dpsl}
In the previous test case, which was a steady state problem, we found that the GPE+BV solver can more effectively damp the artificial acoustic waves. We extend our analysis further to investigate how the addition of bulk viscosity affects the flow field of an unsteady problem. For this purpose, the doubly periodic shear layer (DPSL) on a unit square domain [$1 \times 1$] is chosen~\cite{bell1989,Brown1995,minion1997}. The initial conditions are as follows
\begin{subequations}
	\begin{align}
	u &=  \begin{cases}
	\tanh(\rho (y-0.25))      & \text{if} \hspace{0.2cm} {y \le 0.5} \\
	\tanh(\rho (0.75-y))      & \text{otherwise}
	\end{cases} \\
	v &= \delta \sin (2 \pi (x + 0.25)) \\
        p &= 0.
	\end{align}
\end{subequations}
In the above equation, the value of $\rho$ determines the thickness of the shear layer, and $\delta$ represents the amplitude or strength of the initial perturbation. We set $\rho=80$ and $\delta = 0.05$ and use a uniform grid of $512 \times 512$, which is necessary to avoid spurious vortices~\cite{Shah2010,Toutant2018}. We set $Re=10^{4}$, and the simulation is run until $t=1$, with $\Delta t = 10^{-5}$. 

Due to the initial perturbation, the shear layers roll up into two separate vortical structures. A direct comparison between the vorticity contours at $t=1$ from both the solvers is presented in figure~\ref{dpsl_vortComparison}. For the GPE+BV solver, we use, $\mathcal{B}=50 \Delta x$ (corresponds to $\mathcal{B} Re = 976.5625$). We observe that the contours from both simulations overlap with each other except for the contour $\omega_z = 0$, for which the GPE+BV solver produces a more accurate result. Moreover, the contours in figure~\ref{dpsl_vortComparison} qualitatively resemble the vorticity plots, without any spurious vortices, reported in literature~\cite{Clausen2013, hashimoto2018, Toutant2018}. We stress that, the filled contours from both simulations are indistinguishable. We purposefully included the line contours to highlight such negligible differences.

\begin{figure}[!ht]
	\centering
	\subfigure[]
	{
		\includegraphics[trim = 20mm 0mm 20mm 0mm, clip, width=7.5cm]{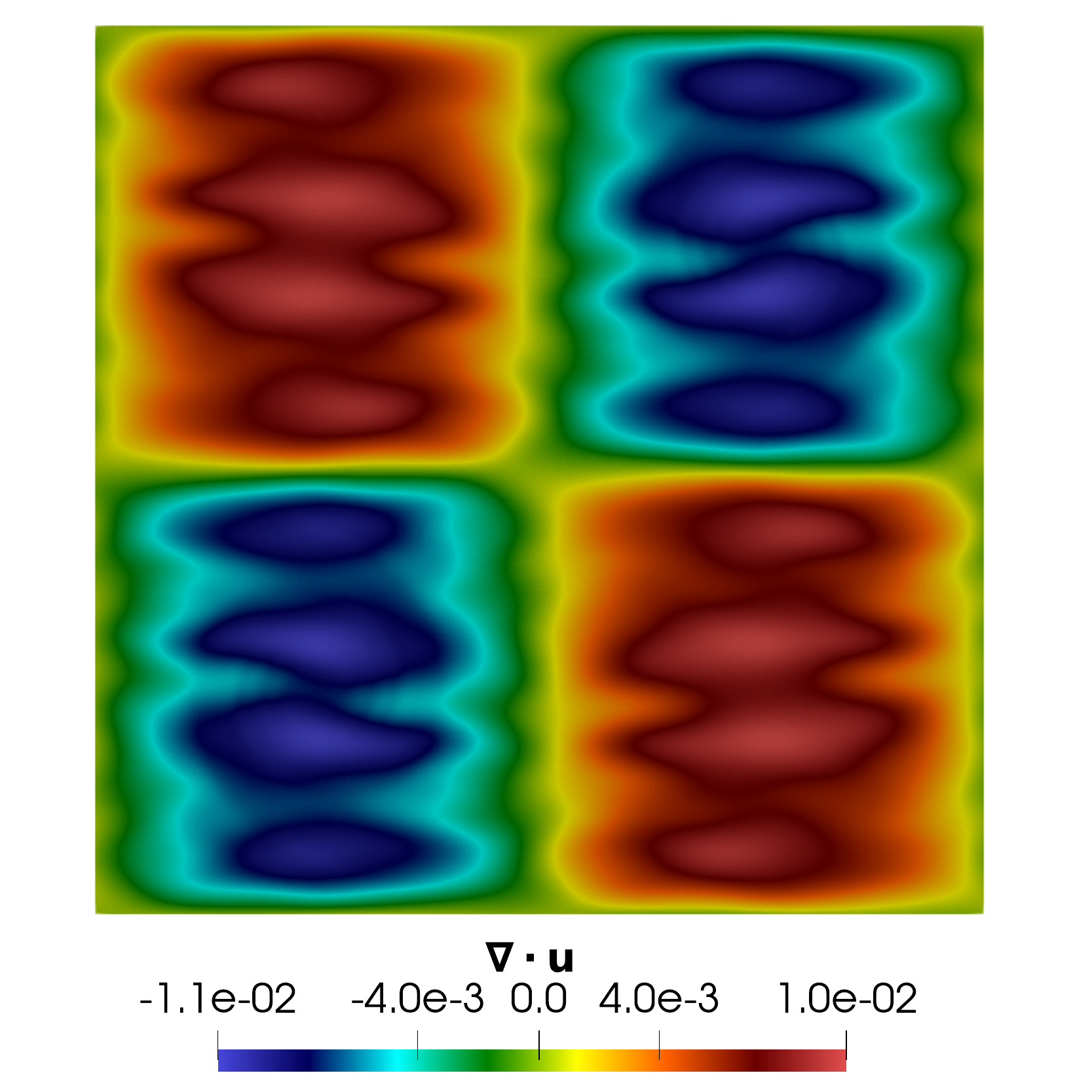}
	}
	\subfigure[]
	{
		\includegraphics[trim = 20mm 0mm 20mm 0mm, clip, width=7.5cm]{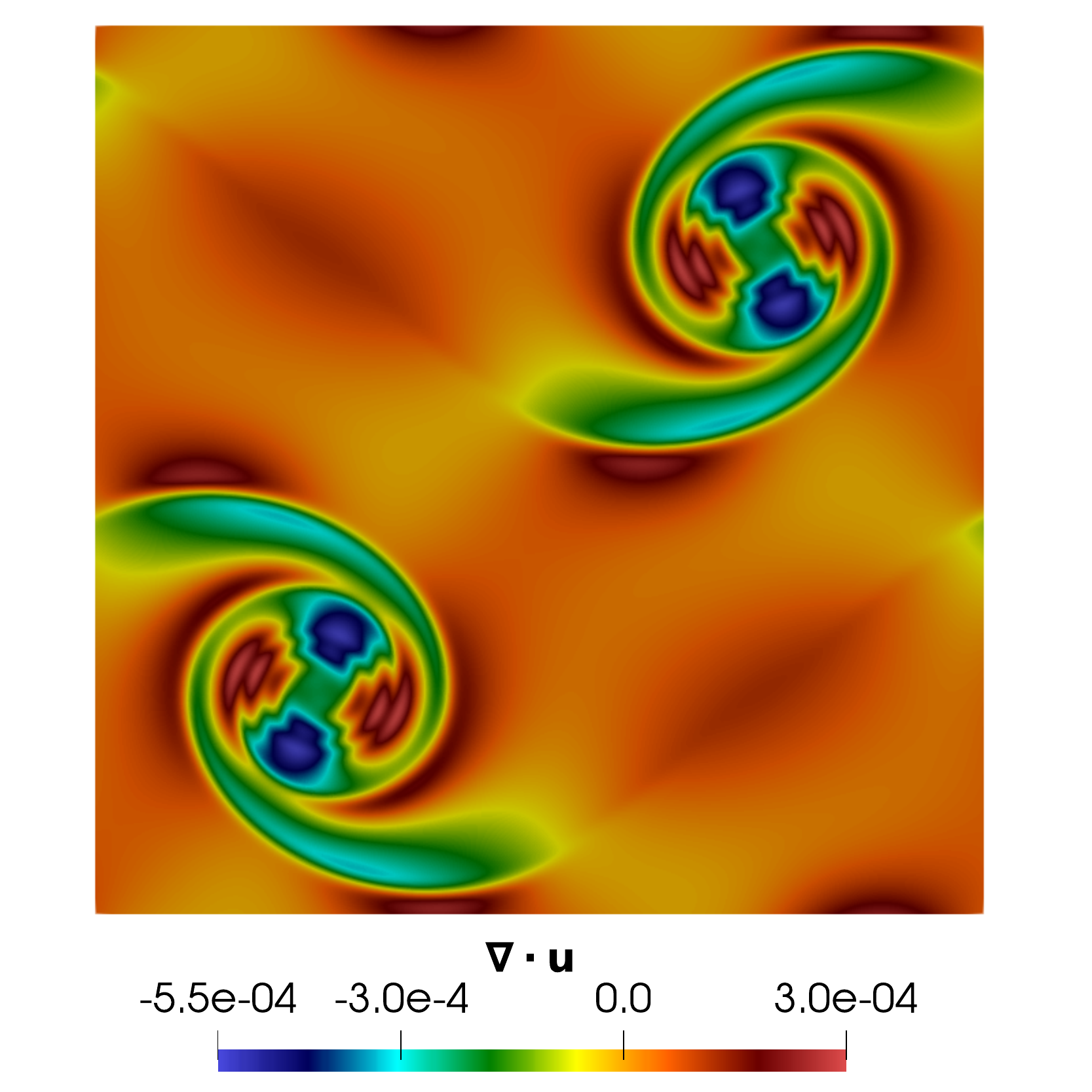}
	}
	\caption{Contours of velocity divergence for the doubly periodic shear layer test case with a grid of $512 \times 512$ at $t=1$ using (a)~GPE solver (b)~GPE+BV solver. }
	\label{dpsl_vel_div}
\end{figure}

The significant advantage of using GPE+BV over GPE can be seen from figure~\ref{dpsl_vel_div}, which illustrates the mass conservation error, quantified by the divergence of the velocity vector, in the whole domain at $t=1$. While GPE produced an error of the order $10^{-2}$, the artificial bulk viscosity present in GPE+BV reduced this error to $10^{-4}$. Thus, a reduction in error of two orders of magnitude is achieved by the addition of $\mathcal{B}$. A major limitation of weakly compressible methods, like GPE, is the large mass conservation error, as these methods don't enforce strict incompressibility. Here, by adding a simple term to the momentum equation and without any complicated modifications, we are able to downscale this error by two orders of magnitude. To the best of our knowledge, such a large reduction in mass conservation error is not reported in the literature.

\begin{figure}[!ht]
	\centering
	\includegraphics[trim = 0mm 0mm 0mm 0mm, clip, width=10cm]{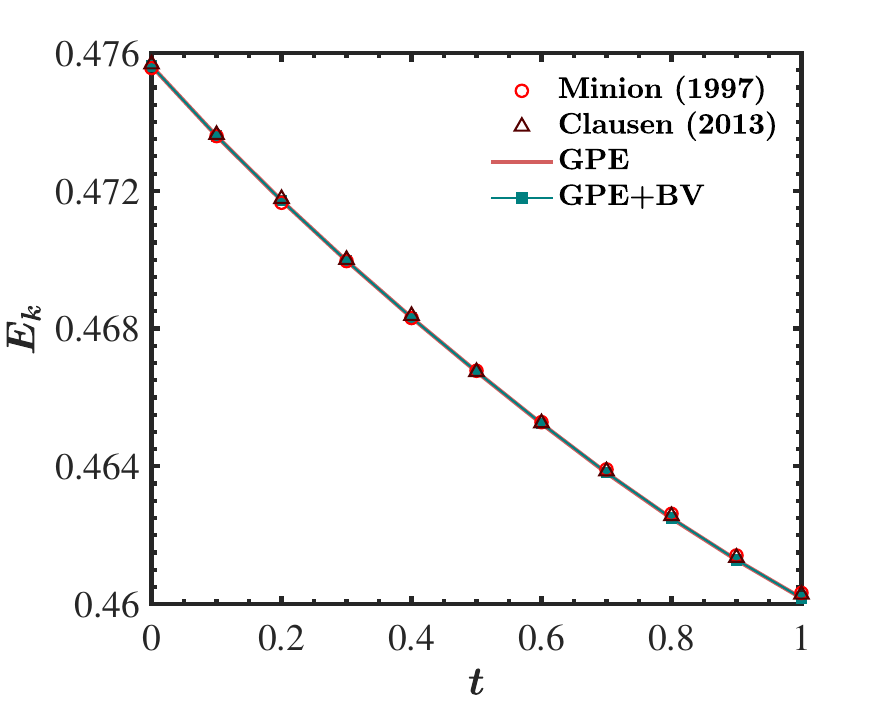}
	\caption{Evolution of average kinetic energy ($E_k$) for the doubly periodic shear layer test case with a grid of $512 \times 512$. The computed values are compared with that reported by Minion and Brown~\cite{minion1997} and Clausen~\cite{Clausen2013}.}
	\label{dpsl_keVsTime}
\end{figure} 

Having demonstrated the reduced mass conservation error, it is also important to verify that the addition of $\mathcal{B}$ does not induce additional error in any other form. In order to check this, we computed the evolution of average kinetic energy,
\begin{equation}
    E_k = \frac{1}{ \volume} \int_\volume  \frac{\left( \mathbf{u} \cdot \mathbf{u}\right)}{2} d\volume,
    \label{ke_equation}
\end{equation}
where $\volume$ corresponds to the total volume of the computational domain. The variation of $E_k$ with time obtained from GPE+BV and reference results are presented in figure~\ref{dpsl_keVsTime}. The exact agreement of results of the GPE+BV solver and the other weakly compressible approach without the bulk viscosity~\cite{Clausen2013}, and from the solution of incompressible Navier-Stokes equation~\cite{minion1997} underlines that the addition of $\mathcal{B}$ does not introduce any other error. This is due to the fact that bulk viscosity directly works with the divergence of velocity field only, leaving all other incompressible components intact~\cite{sun2023}.

The velocity divergence of the weakly compressible approaches cannot be improved with mesh refinement, which was discussed by Toutant~\cite{Toutant2018}. To address this, he rather chose to increase the diffusion of artificial acoustic waves by reducing $Pr$. From his parametric study on $Pr$, the minimum value of velocity divergence achieved was of the order $10^{-3}$ with $Pr=0.01$. However, as given in equation~\eqref{time_step_eqn_Pr}, this mandates a reduction in $\Delta t$. In contrast, with the GPE+BV solver with $Pr=1$, we are able to attain velocity divergence of the order $10^{-4}$ without reducing $\Delta t$. This confirms that the bulk viscosity serves as a computationally more efficient procedure to annihilate artificial acoustic waves when compared to the pressure diffusion term.

The FEM community has also used a similar approach of including the bulk viscosity term~\cite{de_mulder1998,rasthofer2018}, which they call `grad-div' stabilisation. Even for the incompressible flow formulations, it has been reported that such a modification gives a reduced velocity divergence error~\cite{olshanskii2009}.

\subsection{Laminar Taylor-Green vortex problem: Effect on the order of convergence}
\label{tgv}
The Taylor Green vortices (TGV)~\cite{taylor1937} is a widely used test case for verifying the order of accuracy of numerical algorithms. Our objective here is to investigate how the artificial bulk viscosity term affects the order of convergence. The analytical solution for this problem, in a doubly-periodic unit square, is given by the following formula~\cite{Toutant2018}, and the initial conditions are obtained from the same by setting $t=0$. 
\begin{subequations}
	\begin{align}
	u(x, y, t) &= \text{cos}(2\pi x) \text{sin}(2\pi y)e^{-\frac{8 \pi ^2}{Re}t} \\
	v(x, y, t) &= -\text{sin}(2\pi x) \text{cos}(2\pi y)e^{-\frac{8 \pi ^2}{Re}t} \\
	p(x, y, t) &= -\frac{1}{4}(\text{cos}(4\pi x) + \text{cos}(4\pi y))e^{-\frac{16 \pi ^2}{Re}t}
	\end{align}
	\label{tgv_exact}
\end{subequations}

\begin{figure}[!ht]
	\centering
	\subfigure[]
	{
		\includegraphics[trim = 0mm 0mm 0mm 0mm, clip, width=7.5cm]{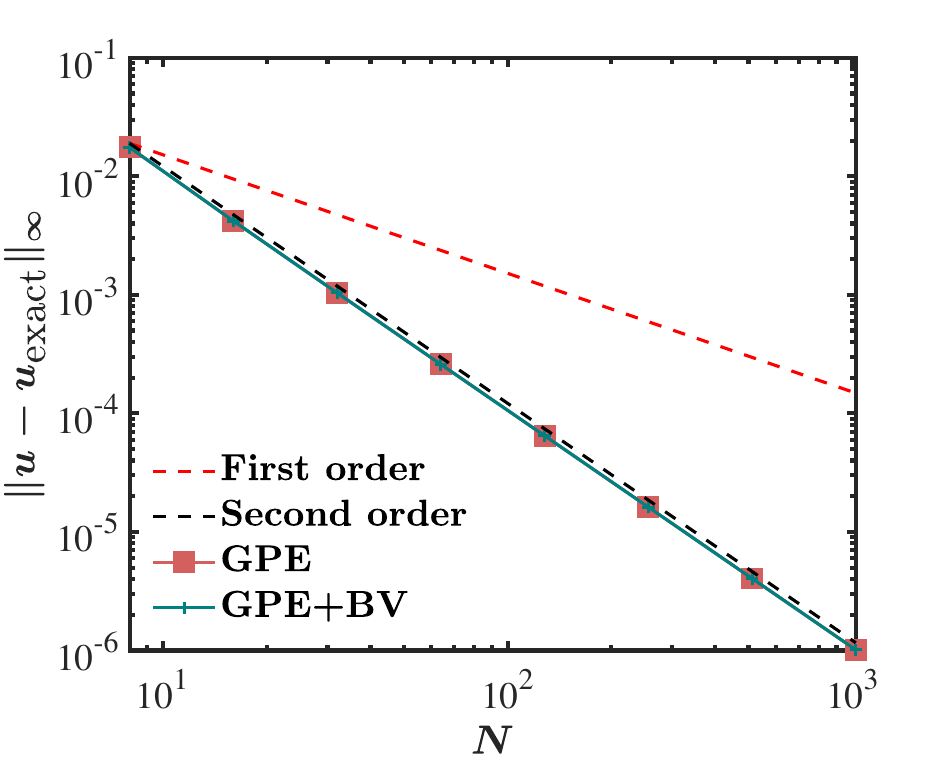}
	}
	\subfigure[]
	{ 
		\includegraphics[trim = 0mm 0mm 0mm 0mm, clip, width=7.5cm]{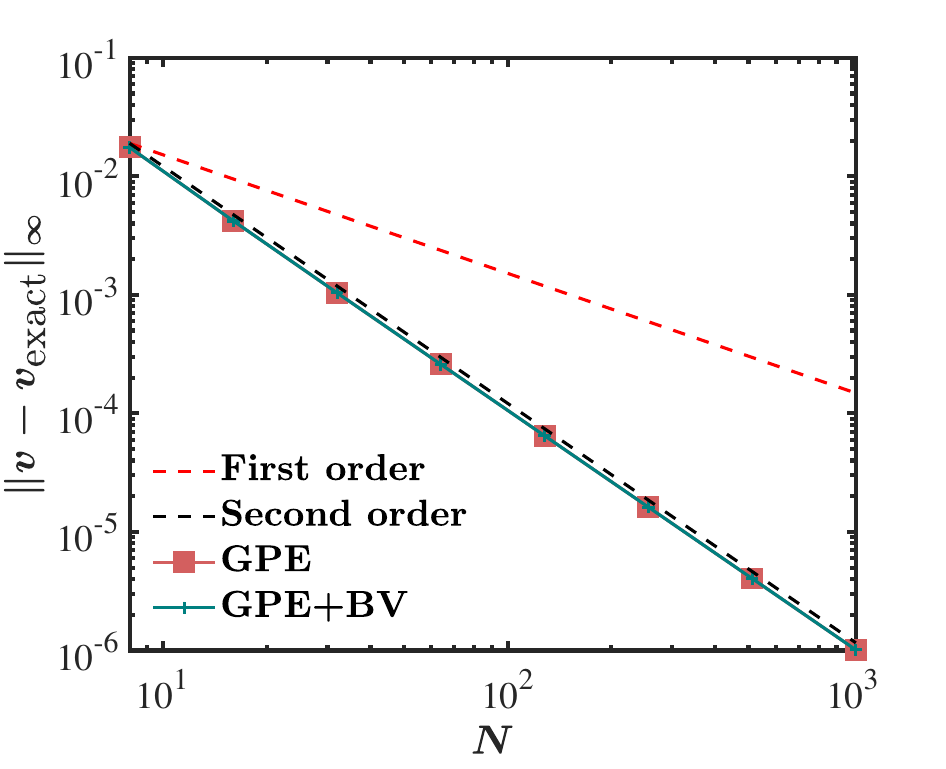}
	}
	\subfigure[]
	{
		\includegraphics[trim = 0mm 0mm 0mm 0mm, clip, width=7.5cm]{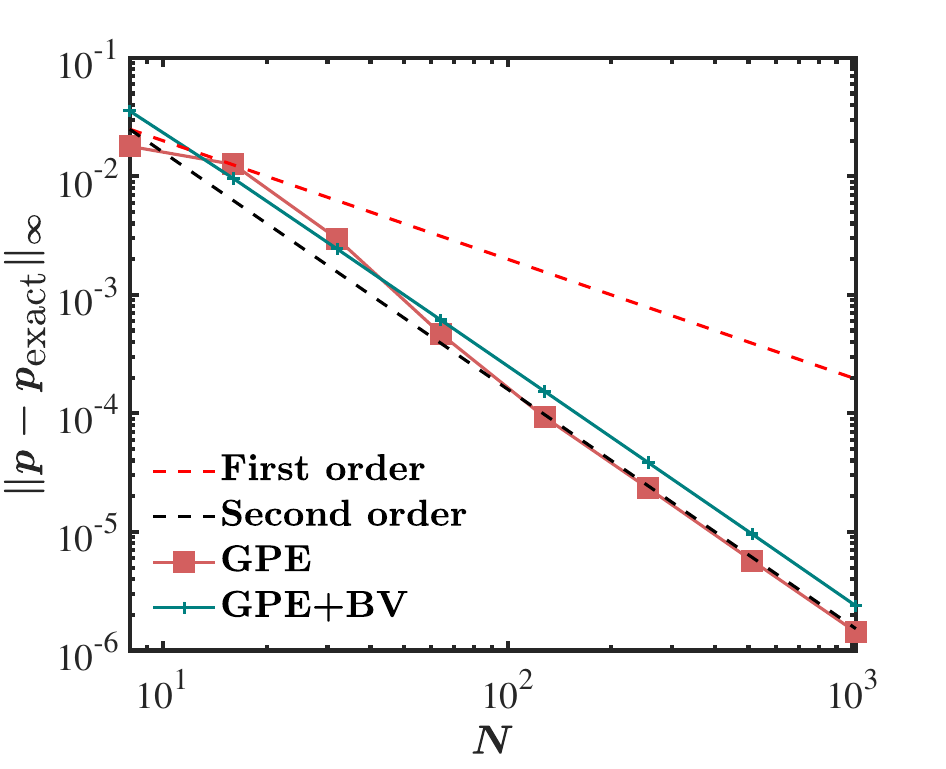}
	}
	
	\caption{Convergence plots for the Taylor-Green vortex problem, with the GPE and GPE+BV solvers (a)~$x-$component of velocity (b)~$y-$component of velocity (c)~pressure.}
	\label{tgv_convergence_plots}
\end{figure}

The TGV case is simulated using GPE and GPE+BV solvers. Similar to the previous test problems, we use a constant value of $\mathcal{B}=50 \Delta x$. We set $Re = 100$ and $\Delta t = 10^{-5}$. 

Convergence plots at $t=1$ for velocity components and pressure are presented in figure~\ref{tgv_convergence_plots}, which shows the variation of $L_\infty$-norm of error with the number of grid points ($N$). For the GPE+BV solver, the convergence study ranges from $\mathcal{B} Re = 19.53125$ for $256 \times 256$ grid to $\mathcal{B} Re = 625.0$ for $8 \times 8$ grid. It can be seen that velocity components exhibit second-order convergence, as expected, for both solvers. However, the order of convergence for pressure exhibits some irregularity for coarser grids when the GPE solver is used. In contrast, we observe uniform second-order convergence for pressure with the GPE+BV solver for the entire range of grid sizes under consideration. 

Toutant~\cite{Toutant2018} also reported the same behaviour for the GPE solver. He deduced the convergence rate of velocity and pressure to be 2 and 1.87, respectively. In comparison, we find that GPE+BV enables a consistent second-order convergence rate for pressure also.

\subsection{Flow over a square cylinder: Time to reach periodic vortex shedding}
\label{cylinder}
The unsteady problems considered in the earlier sections were free-shear flows. In this section, we consider the fluid flow over a square cylinder, an example that involves large-scale boundary layer separation. In this and the next example, the domain is covered with a stretched mesh; thus, the artificial bulk viscosity becomes non-homogeneous and anisotropic as given by equation~\eqref{anisotropic_bulkVisc}. The objective is to investigate the proposed form of $\boldsymbol{\mathcal{B}}$ on (i)~the effectiveness in suppressing the acoustic waves, thereby reaching the periodic shedding state quickly, and (ii)~accuracy of computing integral quantities over a solid surface, such as drag and lift forces.

\begin{figure}[!ht]
	\centering
	\includegraphics[trim = 0mm 0mm 0mm 0mm, clip, width=15cm]{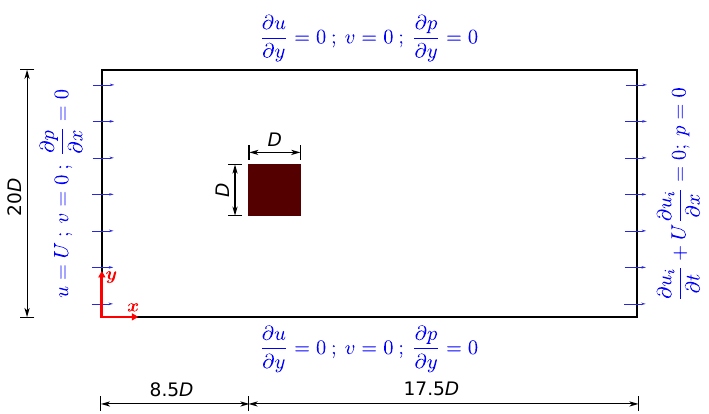}
	\caption{Geometric specifications and boundary conditions of flow over a square cylinder.}
	\label{cylinderSchematic}
\end{figure}

Figure \ref{cylinderSchematic} shows the problem's schematic and boundary conditions. We use a non-uniform mesh of $526 \times 442$, and the complete mesh details are given in table \ref{cylinderMeshTable}. The geometry, boundary conditions and mesh size around the cylinder are the same as that of Sharma and Eswaran~\cite{sharma2004a}. The flow field is initialised to $u=v=p=0$, and the time step of $10^{-4}$ is used to simulate the flow until the periodic state is established. The flow is defined by $Re=100$ and the characteristic velocity, $U=1$. 

\begin{table}[!ht]
	\centering
	\caption{Details of the grid for the case of flow over a square cylinder.}
	\begin{tabular}{|c|cccc|c|cccc|}
		\hline
		\hline
		\multicolumn{5}{|c|}{\textbf{Grid size along $\boldsymbol{x}$ }} & \multicolumn{5}{c|}{\textbf{Grid size along $\boldsymbol{y}$ }}\\
		\hline
		$\boldsymbol{x}$         & 0         & 8.25              & 9.75              & 26       &		$\boldsymbol{y}$     & 0         & 9.25        & 10.75       & 20 \\
  \hline
		$\boldsymbol{\Delta x}$    & 0.25   & 0.008334    & 0.008334       & 0.25  &   $\boldsymbol{\Delta y}$  & 0.25   & 0.008334    & 0.008334    & 0.25\\
		\hline
		\hline
	\end{tabular}
	\label{cylinderMeshTable}
\end{table}

\begin{figure}[!ht]
	\begin{center}
		\centering
		\subfigure[]
		{
			\includegraphics[trim = 0mm 0mm 0mm 0mm, clip, width=15.0cm]{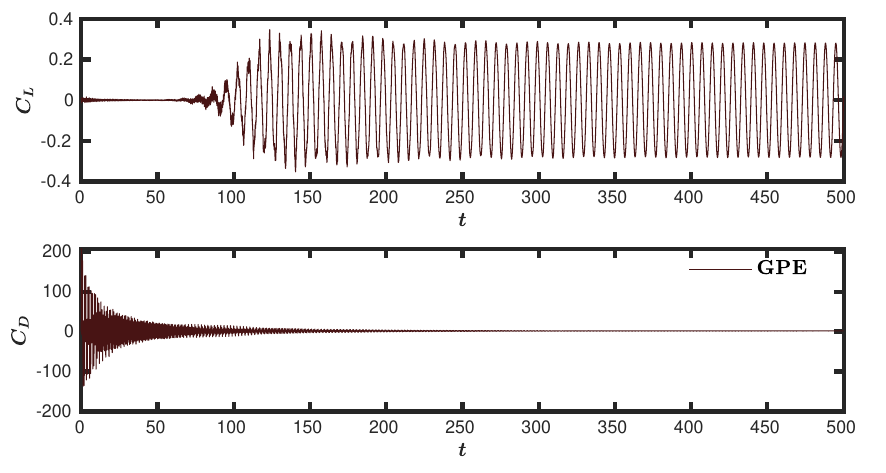}
			\label{cylinder_liftDrag_woBulkVisc_gpe_500}
		}
		\subfigure[]
		{
			\includegraphics[trim = 0mm 0mm 0mm 0mm, clip, width=15.0cm]{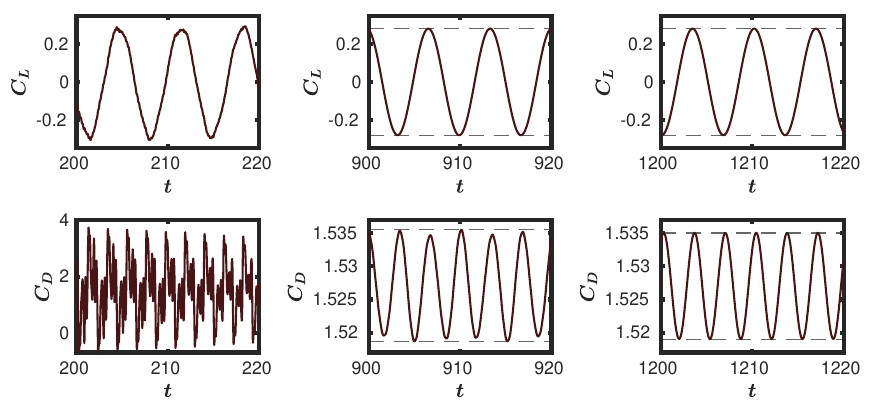}
			\label{cylinder_liftDrag_woBulkVisc_gpe_checkPeriodic}
		}
	\end{center} 
	\caption{The evolution of lift and drag coefficients with respect to time for the flow over a square cylinder problem using the GPE solver ($Pr=1$). (a)~Plots till $t=500$ (b)~Plots at different time intervals to check periodicity.}
	\label{cylinder_liftDrag_gpe}
\end{figure}

We first discuss the results obtained from the GPE solver. After some initial transients, we observe the vortex shedding, which can be graphically visualised from the lift coefficient ($C_L$) and the drag coefficient ($C_D$) plots, shown in figure~\ref{cylinder_liftDrag_woBulkVisc_gpe_500}. The high frequency oscillations present in the plots indicate the acoustic waves visually, and with time they get damped out. To examine whether the periodic state is achieved, we plot $C_L$ and $C_D$ at different time intervals and are presented in figure~\ref{cylinder_liftDrag_woBulkVisc_gpe_checkPeriodic}. We observe that the pressure waves are highly prominent around $t=200$, which gets almost filtered out by $t=900$. However, it is only around $t=1200$ that a complete periodic flow is attained. It can be seen that $C_D$, predominantly contributed by the pressure drag, requires a longer time to stabilise than $C_L$. Thus, the delay in achieving periodic flow is attributed to the slower rate of damping the acoustic pressure oscillations. 

\begin{figure}[!ht]
	\begin{center}
		\centering
			\includegraphics[trim = 0mm 0mm 0mm 0mm, clip, width=15.0cm]{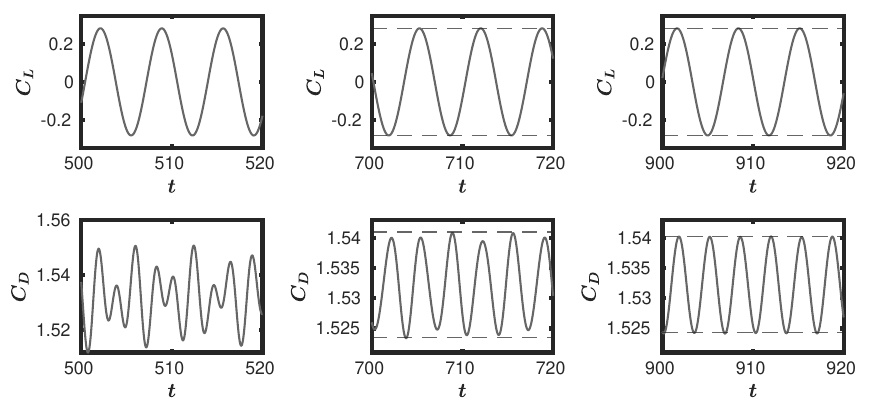}
	\end{center} 
	\caption{The evolution of lift and drag coefficients with respect to time from the simulation of the flow over square cylinder problem using GPE solver ($Pr=0.01$).}
	\label{cylinder_liftDrag_gpe_PrStudy}
\end{figure}

The GPE solver itself has the following mechanisms to dissipate the acoustic waves: viscosity, numerical diffusion, and pressure diffusion. Out of these, only the pressure diffusion term can be specified as an input parameter by adjusting the value of $Pr$ (equation~\eqref{gpe}). Before inspecting the prospects of using the GPE+BV solver, it is important to examine the effectiveness of increasing the pressure diffusion in annihilating the acoustic waves. Accordingly, we simulated the test case with $Pr=0.01$. The corresponding $C_L$ and $C_D$ plots are presented in figure~\ref{cylinder_liftDrag_gpe_PrStudy}. Compared to the results shown in figure~\ref{cylinder_liftDrag_gpe} with $Pr=1$, we observe a slightly faster damping of the pressure oscillations. i.e., the time taken to reach the periodic state is reduced to $t=900$, from $t=1200$ when $Pr=1$. However, this improvement is achieved with a higher computational cost since a lower time-step ($\Delta t = 10^{-5}$) is necessary for stability when $Pr=0.01$. Evidently, faster damping can be achieved with a lower $Pr$ but at a higher computational cost, which is undesirable.

\begin{figure}[!ht]
	\centering
 \subfigure[]
	{
		\includegraphics[trim = 0mm 0mm 0mm 0mm, clip, width=7.5cm]{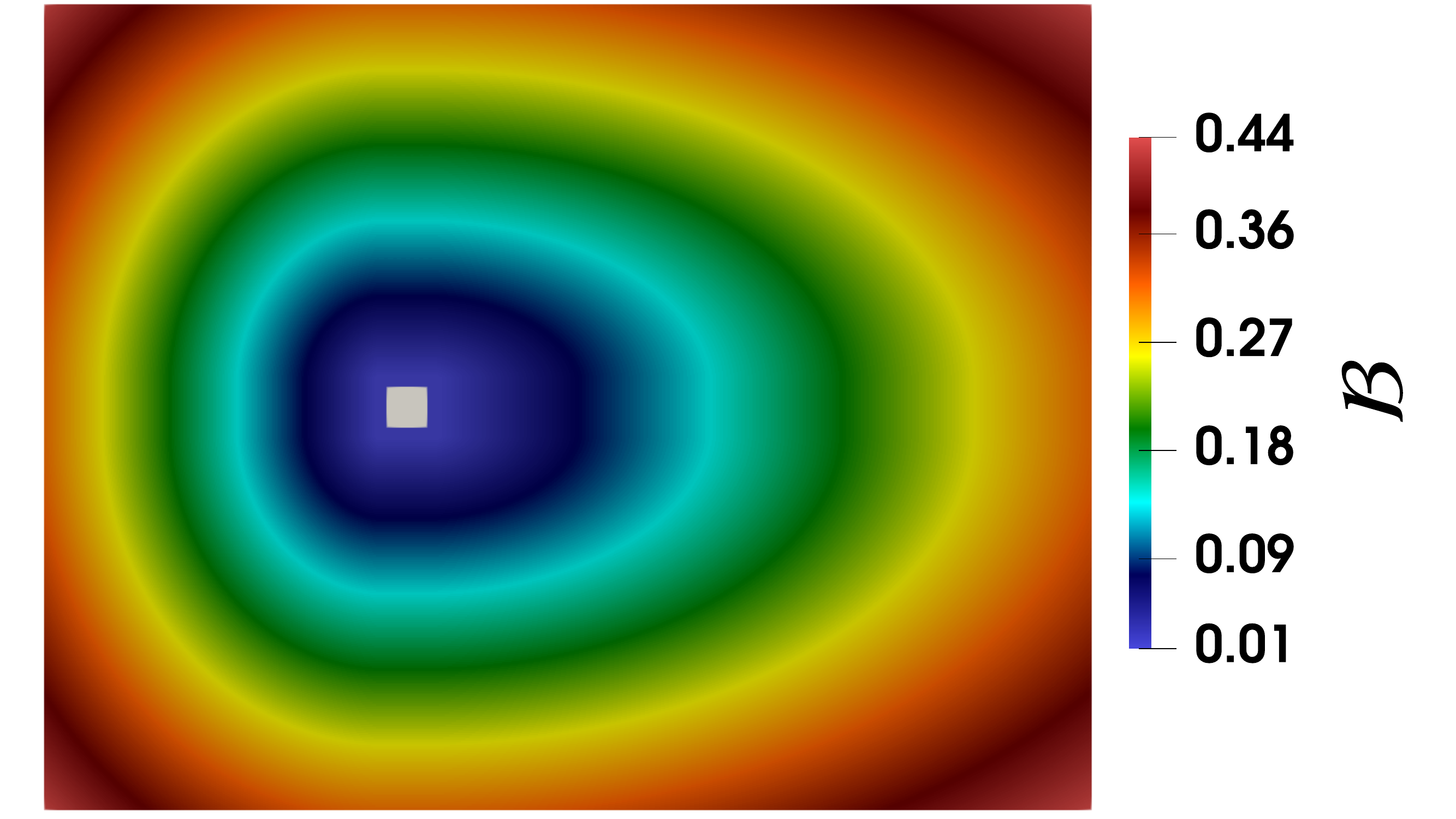}
	} \\
	\subfigure[]
	{
		\includegraphics[trim = 0mm 0mm 0mm 0mm, clip, width=15cm]{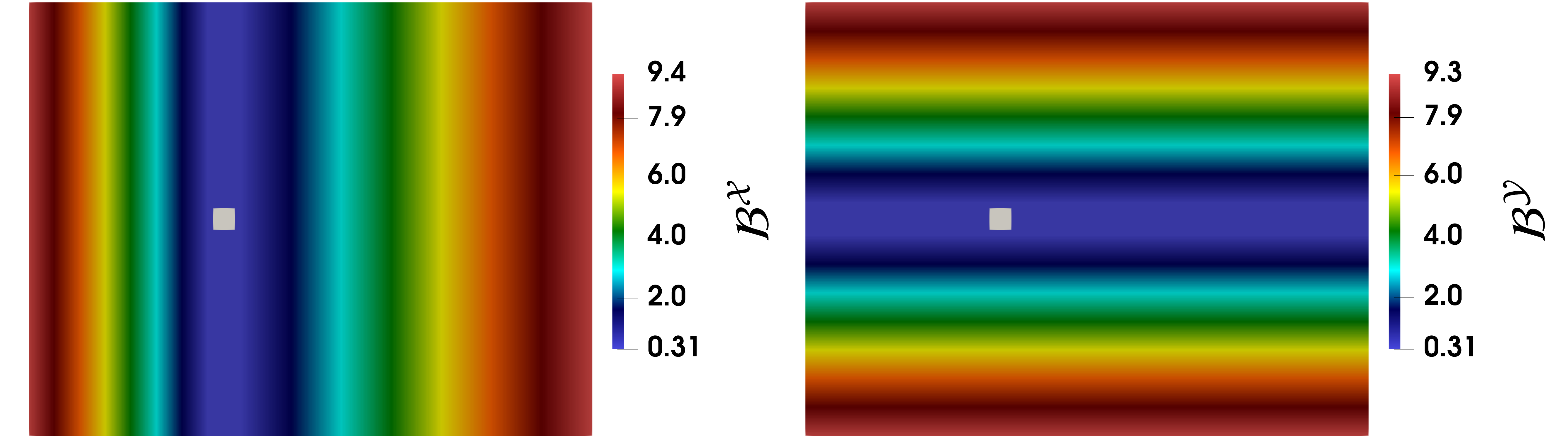}
        }
	\caption{Spatial variation of artificial bulk viscosity coefficient. (a) Non-homogeneous isotropic function, $\mathcal{B} = \Tilde{\lambda} \sqrt {\Delta x^2 + \Delta y^2}$ (b) Components of non-homogeneous anisotropic function $\mathcal{B}^{\mathcal{X}}=\lambda \Delta x$, $\mathcal{B}^{\mathcal{Y}} = \lambda \Delta y$.}
	\label{A_variation}
\end{figure}

Next, we examine the capability of the GPE+BV solver. As discussed earlier, for a non-uniform mesh,  $\boldsymbol{\mathcal{B}}$ is non-homogeneous and anisotropic. In order to isolate the influence of the non-homogeneity and the anisotropic nature, we performed simulations with the following variants of $\boldsymbol{\mathcal{B}}$.
\begin{enumerate}
    \item Homogeneous isotropic function, $\boldsymbol{\mathcal{B}} = \lambda \Delta_{\textrm{min}} \bf{I}$, where $\Delta_{\textrm{min}}$ is the smallest mesh size of the entire domain.  For the considered mesh, $\Delta_{\textrm{min}} = 0.008334$ (refer table~\ref{cylinderMeshTable})
    \item Non-homogeneous isotropic function, $\boldsymbol{\mathcal{B}} = \Tilde{\lambda} \sqrt {\Delta x^2 + \Delta y^2} \bf{I}$, where $\Delta x$ and $\Delta y$ denote the size of the control volume under consideration.
    \item Non-homogeneous anisotropic function, $ \boldsymbol{\mathcal{B}} = 
                                            \begin{bmatrix}
                                                \mathcal{B}^{\mathcal{X}} & 0 \\
                                                0 & \mathcal{B}^{\mathcal{Y}}
                                        \end{bmatrix}$, where,  $\mathcal{B}^{\mathcal{X}}=\lambda \Delta x$ and $\mathcal{B}^{\mathcal{Y}} = \lambda \Delta y$.
\end{enumerate}
We use $\lambda=37.5$, the highest value that can be used without reducing the time step. Also, this corresponds to $\mathcal{B}_{\max} Re = 937.50$ and $\mathcal{B}_{\min} Re = 31.2525$. In order to ensure stability for the variant-2, we set $\tilde{\lambda}=\lambda/AR_{max}$, where $AR_{max}$ is the maximum aspect ratio of the grid. For both the non-homogeneous variants, the distribution of bulk viscosity in the domain is presented in figure~\ref{A_variation}.

\begin{figure}[!ht]
	\begin{center}
		\centering
		\subfigure[]
		{
			\includegraphics[trim = 0mm 0mm 0mm 0mm, clip, width=15.0cm]{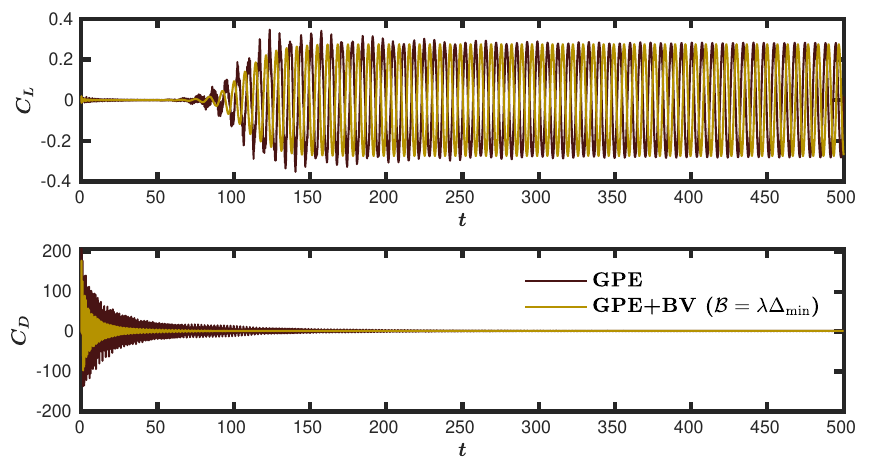}
			\label{cylinder_liftDrag_comparison_lambdaConst_till500s}
		}
		\subfigure[]
		{
			\includegraphics[trim = 0mm 0mm 0mm 0mm, clip, width=15.0cm]{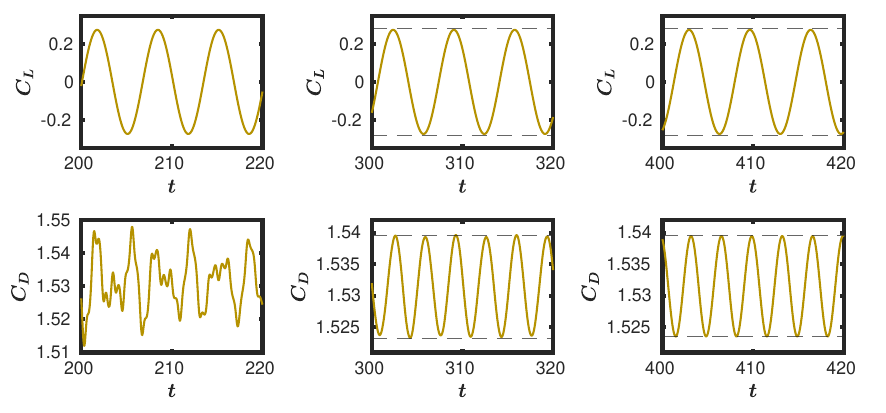}
			\label{cylinder_liftDrag_AAConst_gpe_checkPeriodic}
		}
	\end{center} 
	\caption{Plots of $C_D$ and $C_L$ for the flow over a square cylinder problem using the homogeneous isotropic variant of GPE+BV solver. (a)~Plots till $t=500$ (b)~Plots at different time intervals to check periodicity. }
	\label{cylinder_liftDrag_comparison_lambdaConst}
\end{figure}

The first variant we considered is the homogeneous isotropic function. We observe a drastic reduction in pressure oscillations and faster damping compared to the GPE solver, as evident from figure~\ref{cylinder_liftDrag_comparison_lambdaConst}. The periodic state is achieved at around $t=400$, which is a significant improvement when compared to the GPE. 

\begin{figure}[!ht]
    \centering
    \includegraphics[trim = 0mm 0mm 0mm 0mm, clip, width=15.0cm]{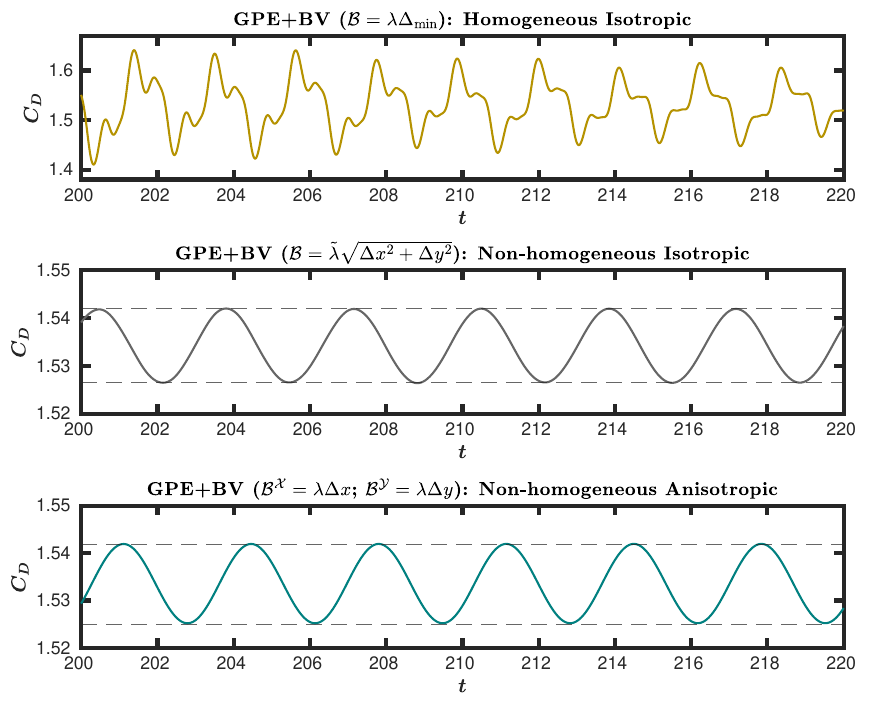}
    \caption{Comparison of $C_D$ for all three variants of $\boldsymbol{\mathcal{B}}$, for the flow over square cylinder problem.}
    \label{cylinder_Dragonly_AFnStudy}
\end{figure}

In figure~\ref{cylinder_Dragonly_AFnStudy}, we compare $C_D$ plots of all three variants between $t=200$ and $t=220$. Both the non-homogeneous isotropic as well as the non-homogeneous anisotropic variants achieve the periodic state in this time frame. This implies significantly more effective damping of the acoustic waves when compared to the isotropic homogeneous version, which is employed by all the existing weakly compressible approaches. 

\begin{figure}[!ht]
    \centering
    \includegraphics[trim = 0mm 0mm 0mm 0mm, clip, width=15.0cm]{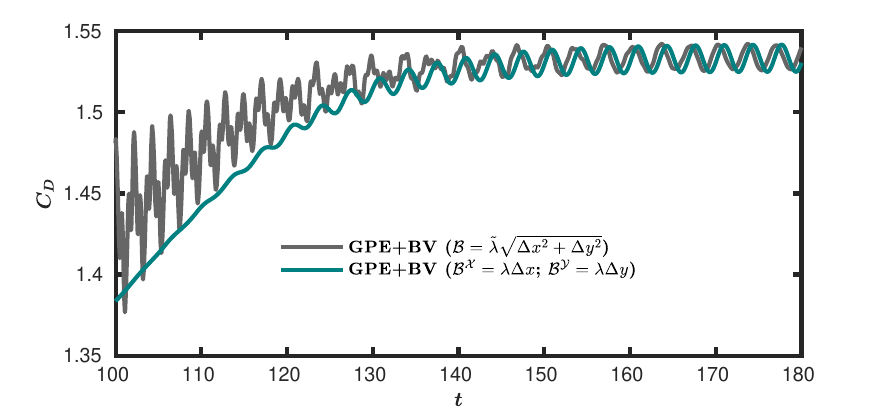}
    \caption{$C_D$ plot for the flow over a square cylinder problem from $t=100$ to $t=180$, using the GPE+BV solver obtained using non-homogeneous isotropic and non-homogeneous anisotropic variants of $\boldsymbol{\mathcal{B}}$. }
    \label{cylinder_Dragonly_AFnStudy_comparison}
\end{figure}

Although variants-2 and 3 achieved periodic states approximately at the same time, inspection of $C_D$ variation during the earlier stages of flow development underlines the superior damping characteristics of the anisotropic variant, as evident from figure~\ref{cylinder_Dragonly_AFnStudy_comparison}. It can be seen that while the non-homogeneous isotropic variant still exhibits unwanted oscillations, the anisotropic variant has almost completely eliminated the pressure acoustics. The consequence of this enhanced performance will be demonstrated for a truly unsteady problem in the next section. 

The computational cost is directly proportional to the time taken to achieve the steady state ($t_{Periodic}$) and the time step. The effect of damping the acoustic waves directly translates into $t_{Periodic}$, and table~\ref{cylinderTimeForPeriodicTable} summarises the same for different variants of $\boldsymbol{\mathcal{B}}$. We normalised the computational costs of all the cases with that incurred for the proposed anisotropic variant. The non-homogeneous variants, which are proposed for the first time in this paper, lead to a reduction of computational cost by at least six times compared to the existing GPE solver. 

\begin{table}[!ht]
	\centering
	\caption{Comparison of time to reach periodic phase using various schemes.}
        \normalsize
	\begin{tabular}{|lccc|}
		\hline
		\hline
		\bf{Scheme}& $\boldsymbol{\Delta t}$  & $\boldsymbol{t_{Periodic}}$        & \bf{Cost}  \\
		\hline
		GPE                    & $10^{-4}$    & 1200   & 6    \\
            GPE+BV ($\boldsymbol{\mathcal{B}} = \lambda \Delta_{\textrm{min}} \bf{I}$)           & $10^{-4}$    & 400    & 2    \\
            GPE+BV ($\boldsymbol{\mathcal{B}} = \Tilde{\lambda} \sqrt{\Delta x^2 + \Delta y^2} \bf{I}$)           & $10^{-4}$    & 200    & 1   \\
            GPE+BV ($\mathcal{B}^{\mathcal{X}}=\lambda \Delta x$ and $\mathcal{B}^{\mathcal{Y}} = \lambda \Delta y$)        & $10^{-4}$    & 200    & 1    \\
		\hline
		\hline
	\end{tabular}
	\label{cylinderTimeForPeriodicTable}
\end{table}

Before concluding, we quantify the accuracy of the proposed GPE+BV solver. For this, we compare the average drag coefficient ($C_{D_{avg}}$), the RMS value of the lift coefficient ($C_{L_{rms}}$) and the Strouhal number ($St$) obtained for all the variants discussed in this section. These values presented in table~\ref{cylinderDataTable} show that all three variants produce accurate results. Specifically, the non-homogeneous anisotropic variant exhibits a deviation of around $2.6\%$, $0.1\%$ and $0.8\%$ respectively for $C_{D_{avg}}$, $C_{L_{rms}}$ and $\textrm{St}$, when compared to Sharma and Eswaran~\cite{sharma2004a}. Thus, with the proposed approach, we achieved sufficient acoustic wave damping without affecting the solver's accuracy.

\begin{table}[!ht]
	\centering
	\caption{Computed values of average drag coefficient ($C_{D_{avg}}$), RMS of lift coefficient ($C_{L_{rms}}$) and Strouhal number ($St$) for the flow over square cylinder problem.}
        \normalsize
	\begin{tabular}{|llll|}
		\hline
		\hline
		$\mathbf{Reference}$& $\boldsymbol{C_{D_{avg}}}$  & $\boldsymbol{C_{L_{rms}}}$        & $\boldsymbol{St}$  \\
		\hline
		Sohankar~\cite{sohankar1998}            & 1.4770  & 0.1560      & 0.1460    \\
		Sharma and Eswaran~\cite{sharma2004a}   & 1.4936  & 0.1922      & 0.1488    \\
		Sen et al.~\cite{sen2011}              & 1.5287  & 0.1928      & 0.1452    \\
		Zhao et al.~\cite{zhao2013}            & 1.47    & 0.156       & 0.146    \\
		Saha and Shrivastava~\cite{saha2015}            & 1.524    & -       & 0.163    \\
		GPE ($Pr=1$)        & 1.5270  & 0.1986      & 0.1480    \\
		GPE ($Pr=0.01$)        & 1.5322 & 0.1997      & 0.1480    \\
             GPE+BV ($\boldsymbol{\mathcal{B}} = \lambda \Delta_{\textrm{min}} \bf{I}$)        & 1.5311  & 0.1948      & 0.15    \\
             GPE+BV ($\boldsymbol{\mathcal{B}} = \Tilde{\lambda} \sqrt{\Delta x^2 + \Delta y^2} \bf{I}$)         & 1.5341  & 0.1935      & 0.15    \\
            GPE+BV ($\mathcal{B}^{\mathcal{X}}=\lambda \Delta x$ and $\mathcal{B}^{\mathcal{Y}} = \lambda \Delta y$)          & 1.5336  & 0.1924      & 0.15    \\
		\hline
		\hline
	\end{tabular}
	\label{cylinderDataTable}
\end{table}

\subsection{Flow over an impulsively started thin plate: Effectiveness in a truly unsteady problem}
\label{impulsePlate}
The previous case was focused on the periodic vortex shedding behind a cylinder, which occurs far away from the initial time. Hence, the effect of artificial acoustic waves in the initial transients did not play a role in the desired periodic state, as can be interpreted from figure~\ref{cylinder_Dragonly_AFnStudy_comparison}. In this section, we investigate the evolution of twin vortices behind an impulsively started thin plate. Our focus is to study in detail the initial stages of development of the flow field and the rapidly varying force coefficients. This is a highly challenging test case for weakly compressible approaches, and we will show that the anisotropic nature of bulk viscosity proposed in this paper provides the best possible results.
\begin{figure}[!ht]
	\centering
		\includegraphics[trim = 0mm 0mm 0mm 0mm, clip, width=15cm]{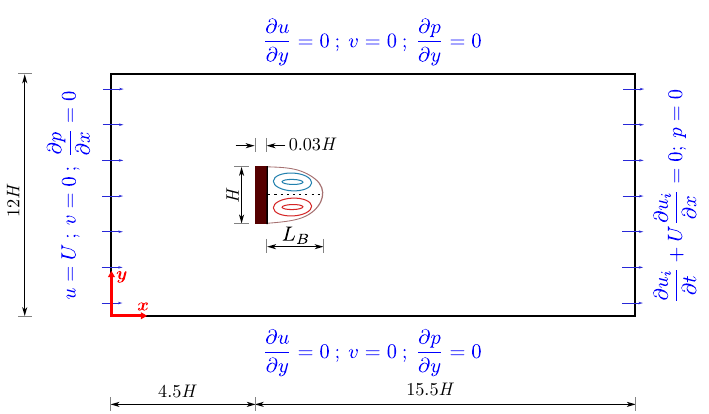}
	\caption{Geometric specifications and boundary conditions of flow over an impulsively started plate.}
	\label{thinPlateSchematic}
\end{figure}
Figure \ref{thinPlateSchematic} shows the schematic of the problem, which is adapted from Kiris and Kwak~\cite{Kiris2002}. The $x$-component of velocity, $u$, is initialized with a constant value of $U=1$ throughout the flow domain to enforce the impulsive starting, and the simulation is carried out at $Re=126$. We use a stretched grid of $679 \times 922$, and the complete mesh details are given in table \ref{acclPlateMeshTable}. The mesh is sufficiently finer around the thin plate to capture the vortex structures properly. The problem is simulated till time $t=10$ with $\Delta t = 10^{-5}$. In order to validate the results from our solver, we simulated this example using the popular open-source software OpenFOAM (denoted as INS in all the plots), which solves the incompressible Navier-Stokes equations. This serves as a reliable reference since the artificial acoustic effects are absent in the `INS' results.
\begin{table}[!ht]
	\centering
	\caption{Details of the grid for the case of flow over an impulsively started thin plate.}
        \footnotesize
	\begin{tabular}{|c|cccccc|c|cccccc|}
		\hline
		\hline
		  \multicolumn{7}{|c|}{\textbf{Mesh size along $\boldsymbol{x}$}} & \multicolumn{7}{c|}{\textbf{Mesh size along $\boldsymbol{y}$}} \\
		\hline
		$\boldsymbol{x}$  & 0      & 3.75	      & 4.25	    & 4.78	    & 5.53	 & 20 & $\boldsymbol{y}$  & 0	    & 4.5	      & 5.25  	    & 6.75	    & 7.5	  & 12 \\
            \hline
		$\boldsymbol{\Delta x}$ & 0.1	& 0.025	      & 0.0025	    & 0.0025	& 0.025	 & 0.1 &	$\boldsymbol{\Delta y}$ & 0.1	& 0.025	      & 0.0025	    & 0.0025	& 0.025	  & 0.1 \\
		\hline
		\hline
	\end{tabular}
	\label{acclPlateMeshTable}
\end{table}

\begin{figure}[!ht]
	\centering
	\subfigure[]
	{
		\includegraphics[trim = 0mm 0mm 11mm 5mm, clip, width=7.5cm]{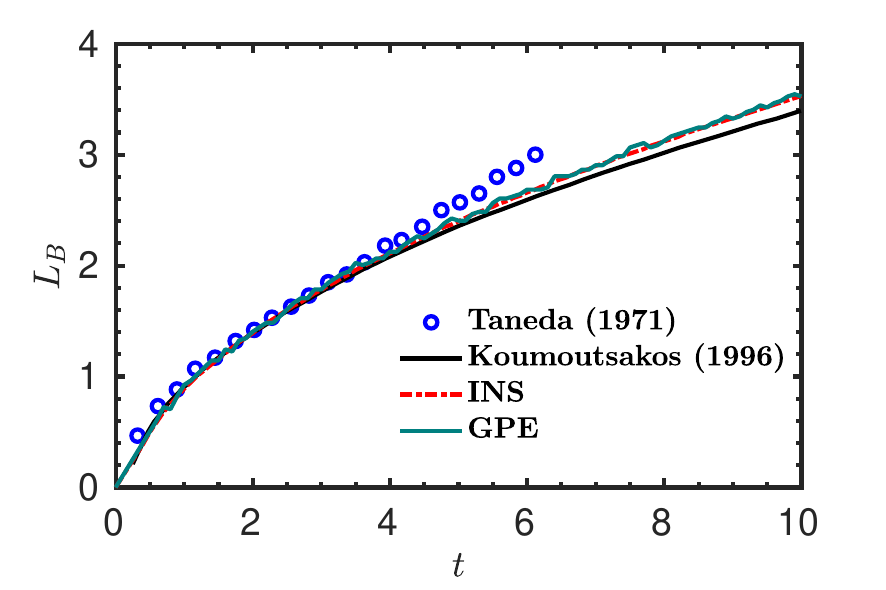}
		\label{thinPlate_bubbleLength_withoutBulkVisc}
	}
	\subfigure[]
	{
		\includegraphics[trim = 0mm 0mm 11mm 5mm, clip, width=7.5cm]{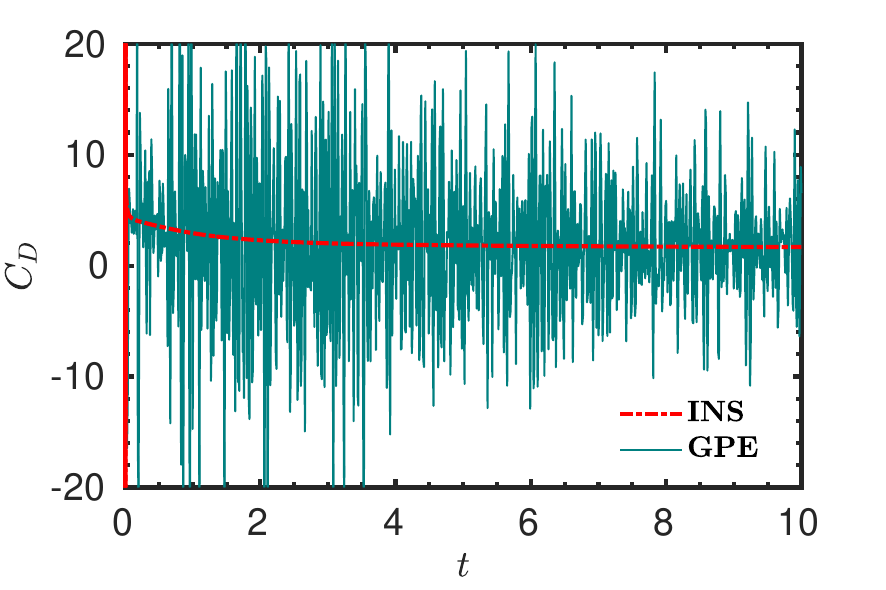}
		\label{thinPlate_drag_withoutBulkVisc}
	}
	\caption{Evolution of (a)~bubble length ($L_B$) compared with~\cite{taneda1971,koumoutsakos1996}; and (b)~drag coefficient ($C_D$) for the flow over an impulsively started thin plate case, using the GPE solver. }
\end{figure}

Firstly, we simulated the test case with the GPE solver. The plot of bubble length ($L_B$) with respect to time is presented in figure~\ref{thinPlate_bubbleLength_withoutBulkVisc}. Even though the GPE solver is able to predict $L_B$ well, the plot exhibits minor fluctuations. This is due to the acoustic waves appearing in the flow field, which induces drastic oscillations in the drag coefficient as shown in figure~\ref{thinPlate_drag_withoutBulkVisc}. As mentioned in \S~\ref{intro}, we can see that the artificial sound waves cause spurious oscillations in $C_D$ to the extent that even its overall variation is untraceable. 

Similar to the previous test cases, reducing $Pr$ leads to suppression of acoustic waves, but at the cost of increased computational time and reduced accuracy. These results are summarised in~\ref{appendix_PrStudy}.

\begin{figure}[!ht]
	\centering
	\subfigure[]
	{
		\includegraphics[trim = 5mm 0mm 11mm 5mm, clip, width=7.5cm]{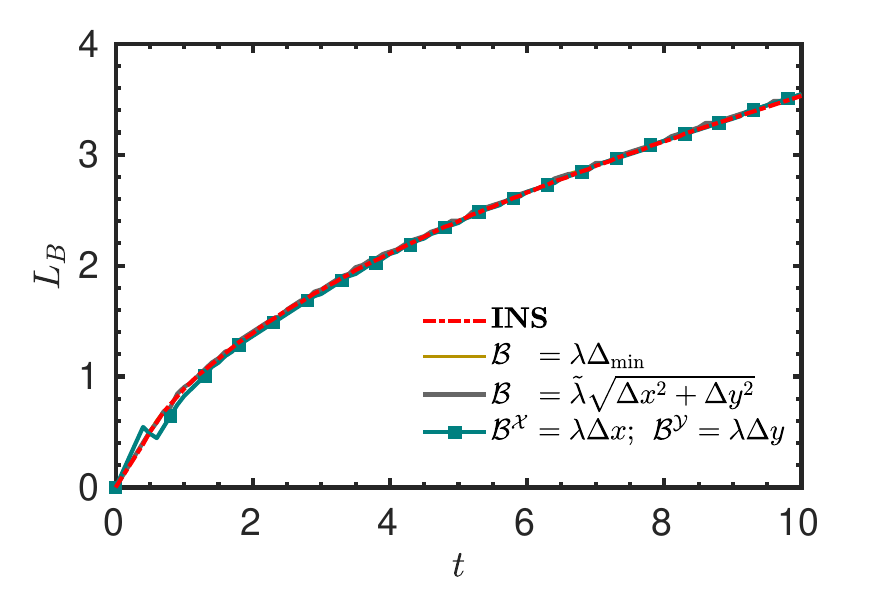}
	}
	\subfigure[]
	{
		\includegraphics[trim = 5mm 0mm 11mm 5mm, clip, width=7.5cm]{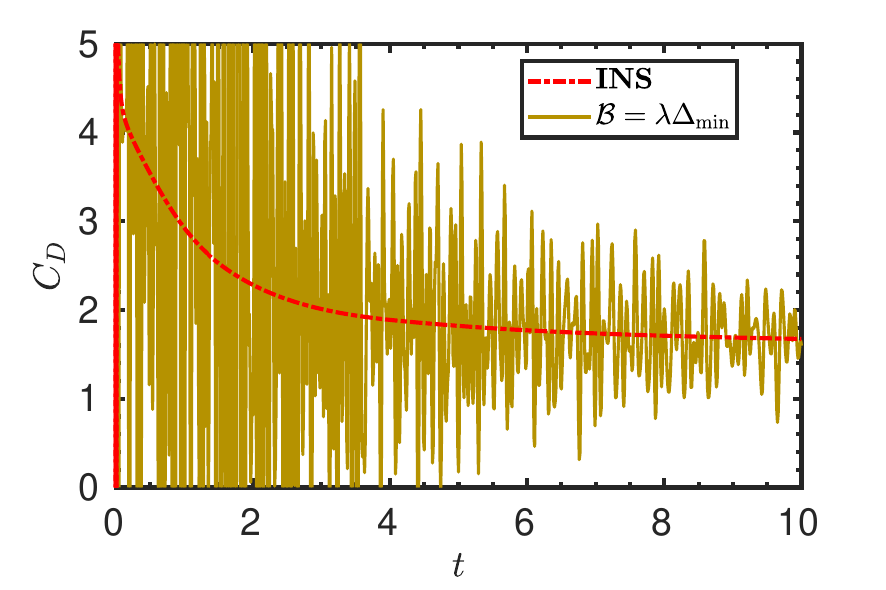}
	}
        \subfigure[]
	{
		\includegraphics[trim = 5mm 0mm 11mm 5mm, clip, width=7.5cm]{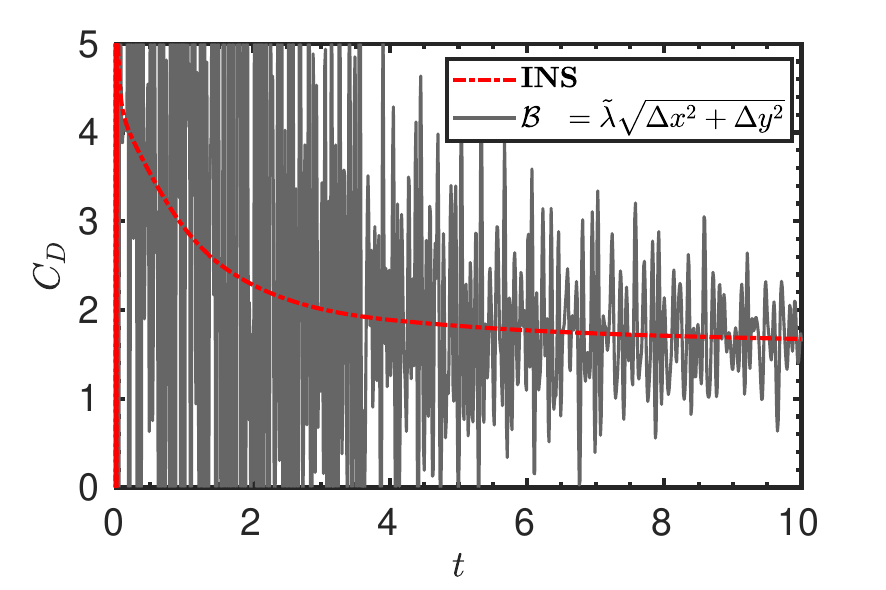}
	}
	\subfigure[]
	{
		\includegraphics[trim = 5mm 0mm 11mm 5mm, clip, width=7.5cm]{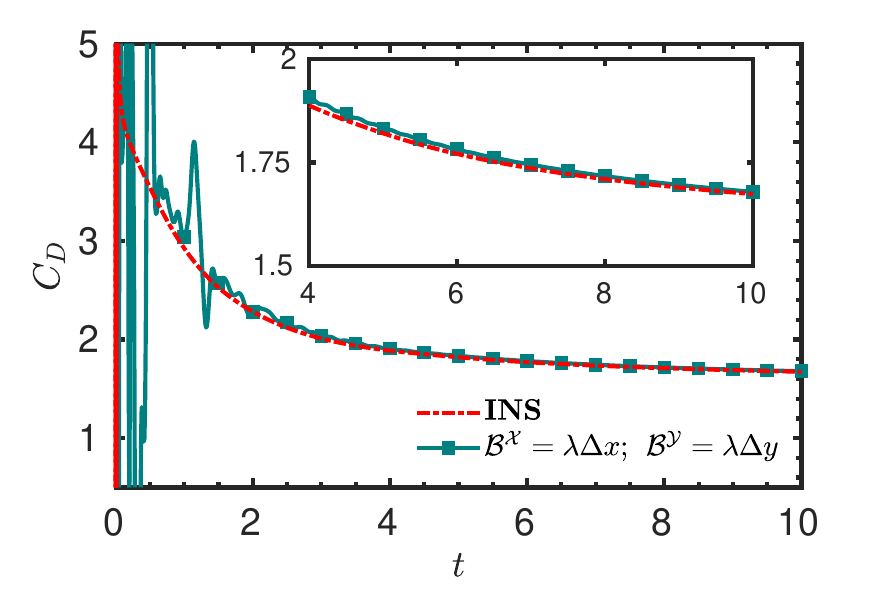}
            \label{acclPlate_CdPlot_linearAnisotropic}
	}
	\caption{Evolution of (a)~bubble length; and drag coefficient for the flow over an impulsively started thin plate case using the GPE+BV solver with different variants of bulk viscosity coefficient ($\boldsymbol{\mathcal{B}}$) such as (b)~homogeneous isotropic (c)~non-homogeneous isotropic (d)~non-homogeneous anisotropic.}
	\label{fn_study}
\end{figure}

Next, we simulate this test case using all three variants of bulk viscosity presented in the previous section and the results are presented in figure~\ref{fn_study}. For all the variants, we use the constant $\lambda= 79$, which indicates $\mathcal{B}_{\max} Re = 995.40$ and $\mathcal{B}_{\min} Re = 24.885$. The following observations can be made from the figure.
\begin{itemize}
    \item All three variants accurately capture $L_B$ as can be seen from figure~\ref{fn_study}(a).
    \item Both the homogeneous isotropic (figure~\ref{fn_study}(b)) and non-homogeneous isotropic(figure~\ref{fn_study}(c)) are ineffective.
    \item The non-homogeneous anisotropic variant of $\boldsymbol{\mathcal{B}}$ completely eliminated the artificial acoustic waves after $t=2$ as is evident from figure~\ref{fn_study}(d).
\end{itemize}
We stress that although for the previous test case, the non-homogeneous isotropic variant produced the same periodic state as that of the non-homogeneous anisotropic case, to capture the dynamics in the early stages of flow development for this truly unsteady problem, the anisotropic nature of bulk viscosity is crucial. This is consistent with the results reported in figure~\ref{cylinder_Dragonly_AFnStudy_comparison}.

\begin{figure}[!ht]
	\centering
    \subfigure[]
    {
	\includegraphics[trim = 0mm 0mm 0mm 0mm, clip, width=7.5cm]{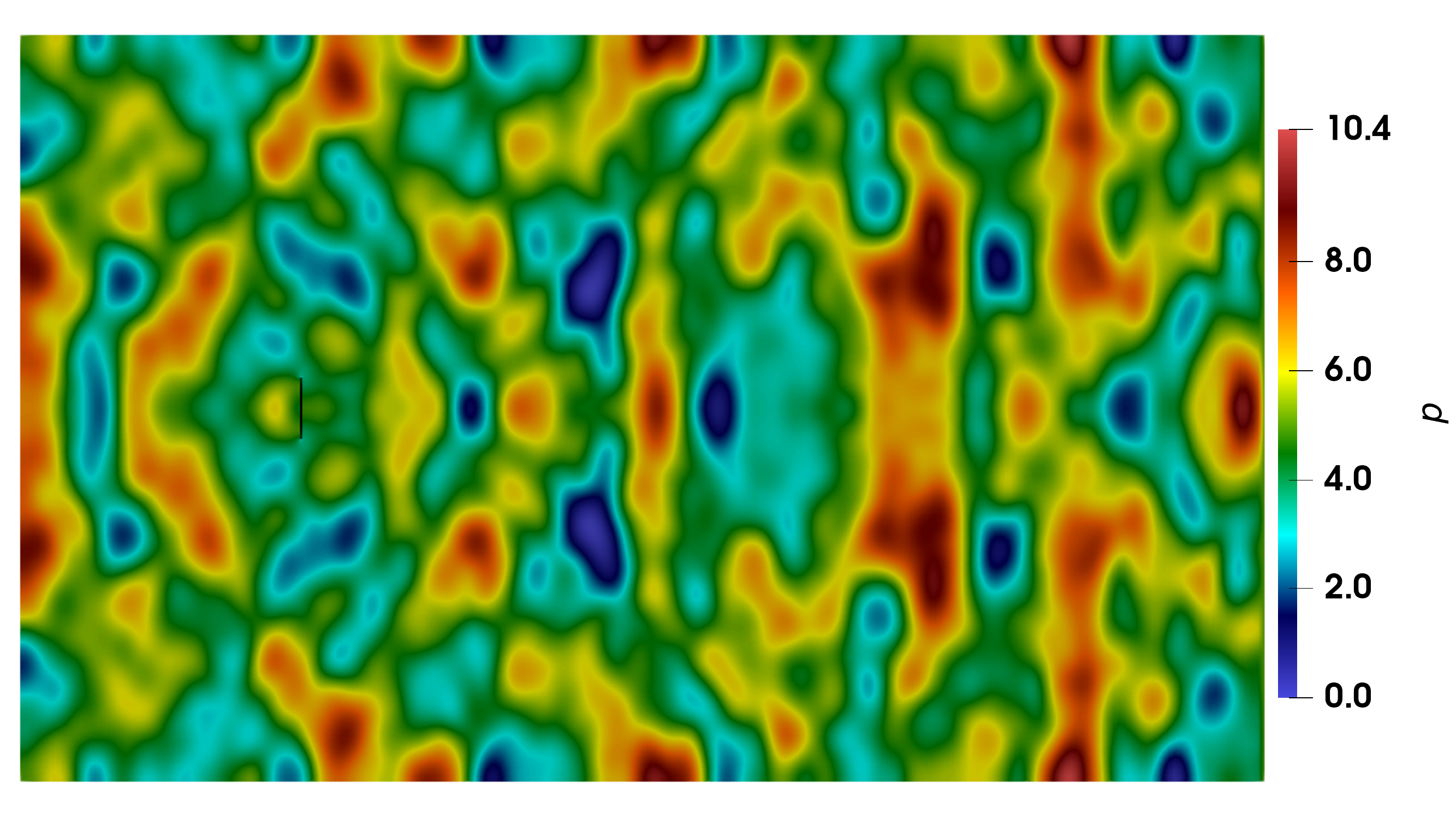}
    }
     \subfigure[]
    {
	\includegraphics[trim = 0mm 0mm 0mm 0mm, clip, width=7.5cm]{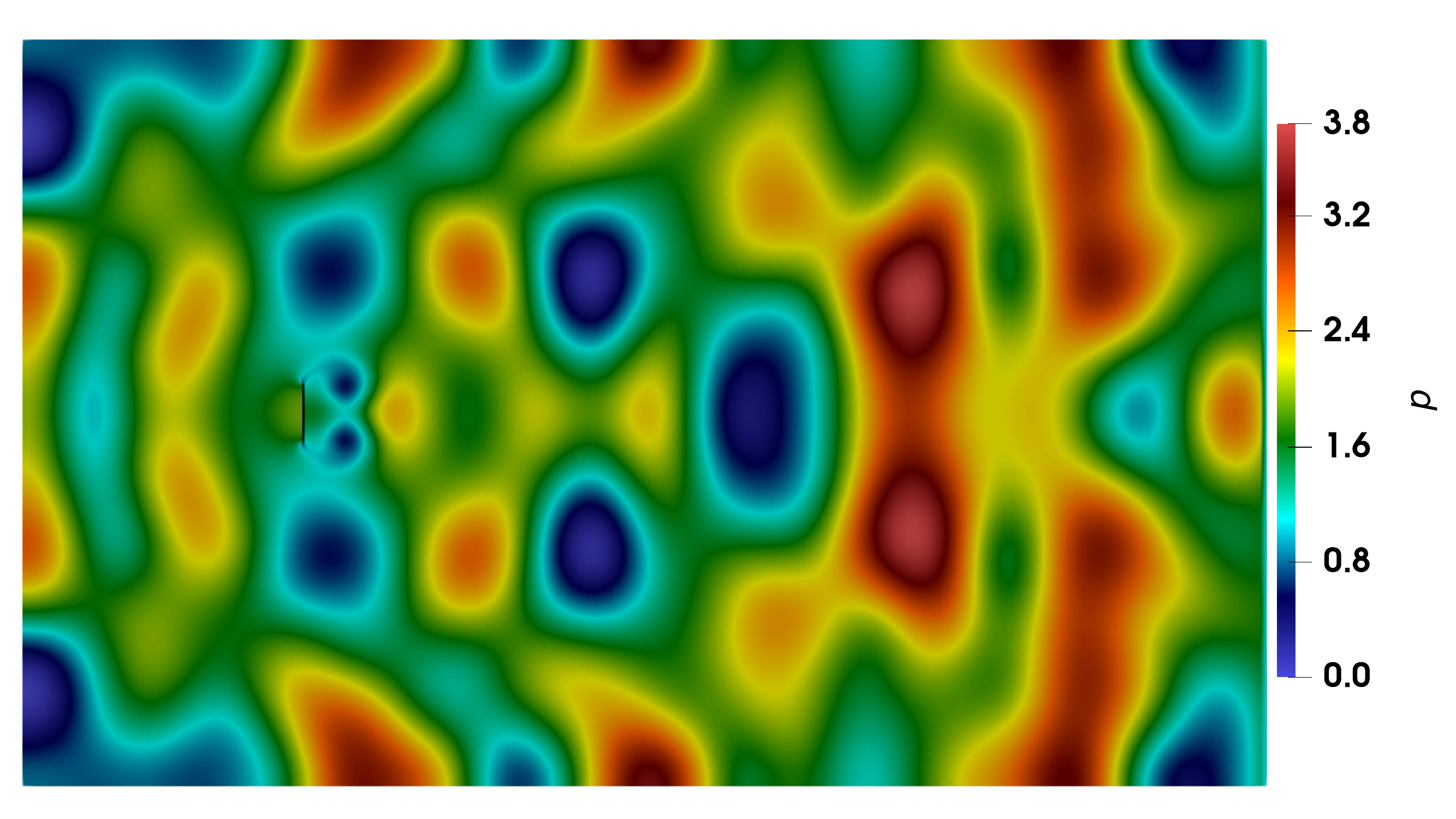}
    }
     \subfigure[]
    {
	\includegraphics[trim = 0mm 0mm 0mm 0mm, clip, width=7.5cm]{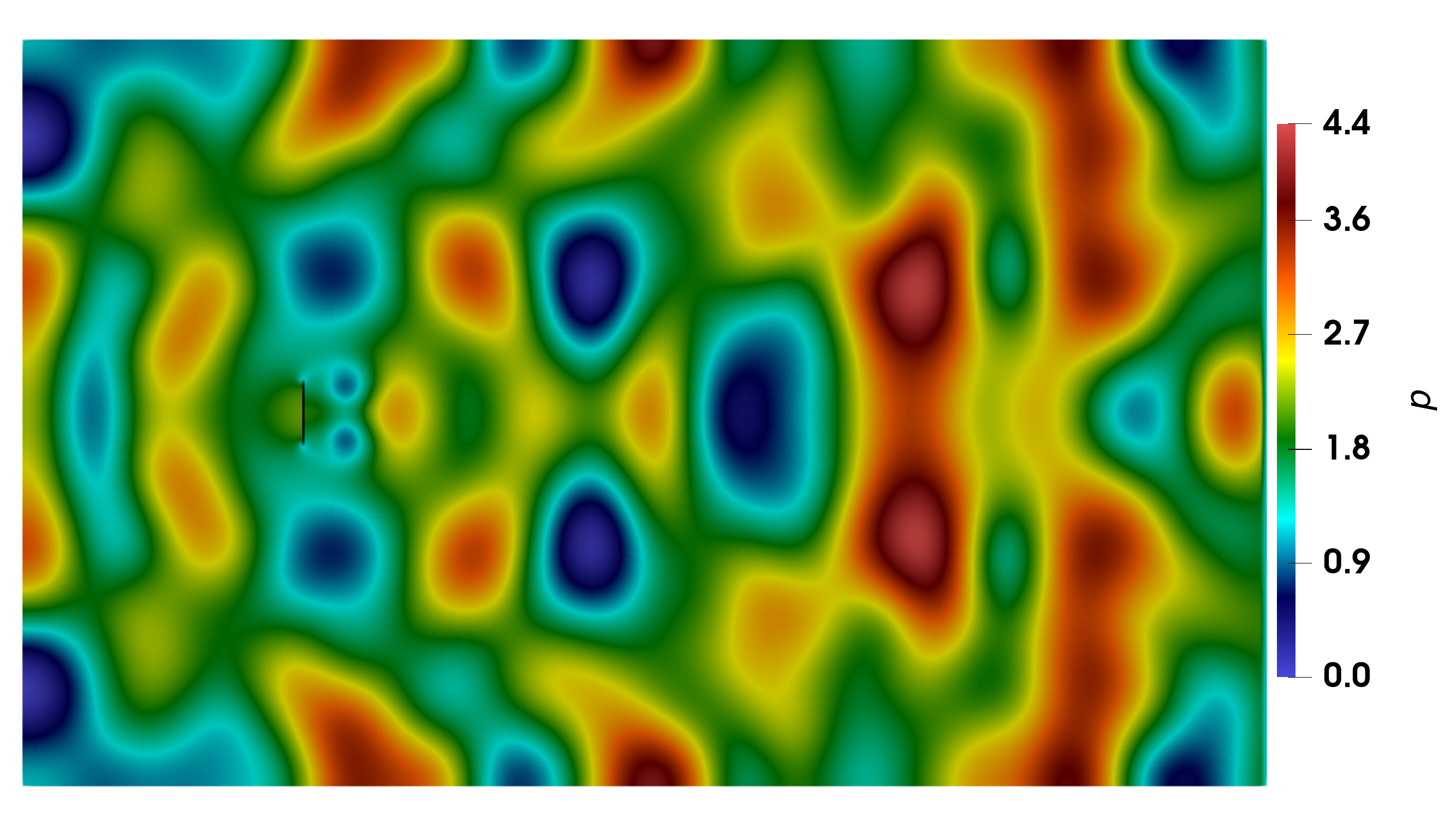}
    }
    \subfigure[]
    {
	\includegraphics[trim = 0mm 0mm 0mm 0mm, clip, width=7.5cm]{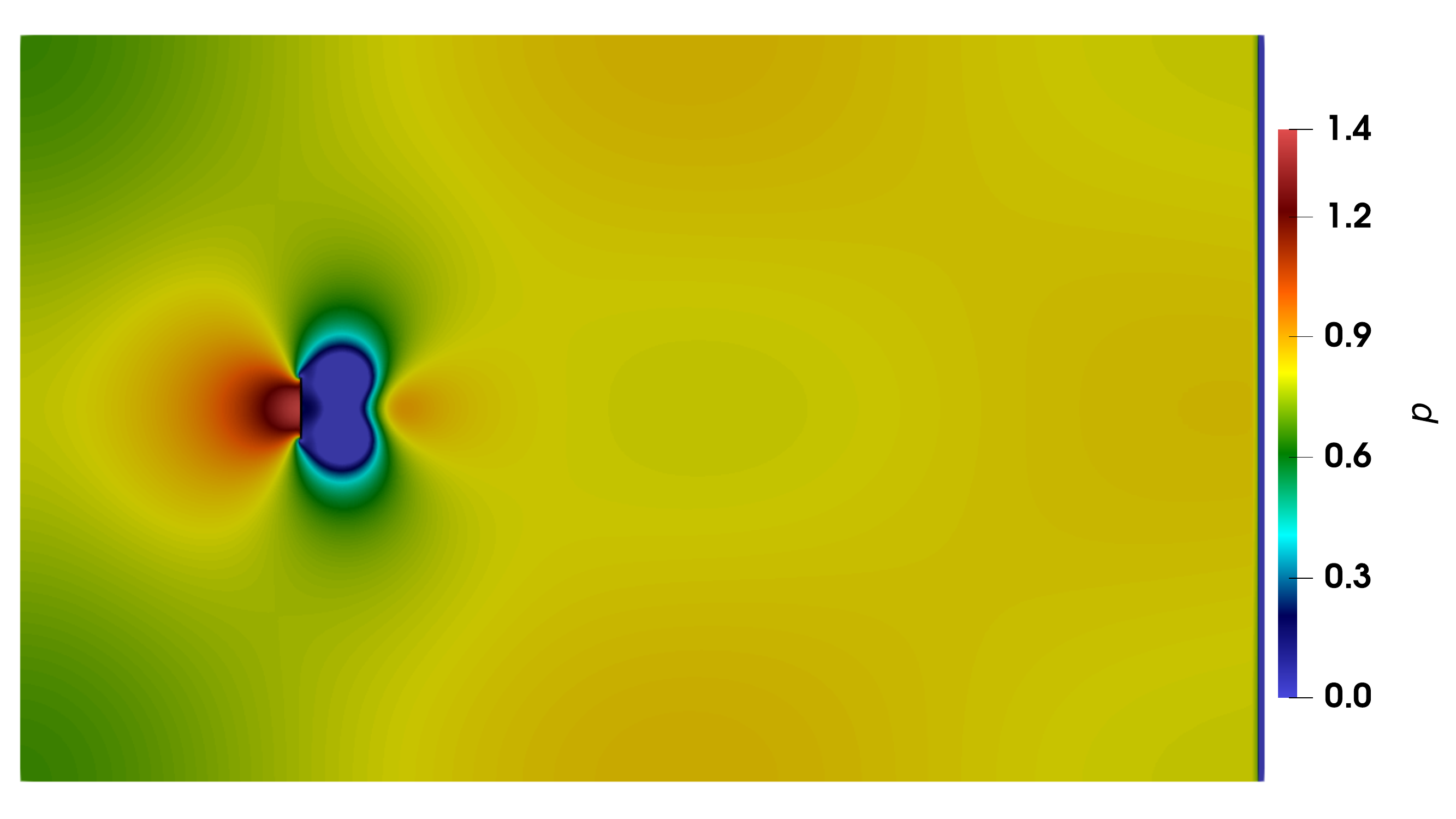}
	\label{thinPlate_A79_p_2s}
    }
    \caption{Pressure field at $t=2$ for the flow over an impulsively started thin plate test case using (a) GPE solver; and the GPE+BV solver with different variants of bulk viscosity coefficient ($\boldsymbol{\mathcal{B}}$) such as (b) homogeneous isotropic (c) non-homogeneous isotropic (d) non-homogeneous anisotropic.}
    \label{thinPlate_p_2s}
\end{figure}

The pressure contours at $t=2$ presented in figure~\ref{thinPlate_p_2s} provide visual evidence for the enhanced performance of the anisotropic variant of bulk viscosity. Consistent with figure~\ref{fn_study}, while the anisotropic variant fully suppressed the acoustic waves, the signature of these waves is clearly seen for GPE and the other isotropic variants. 

From figure~\ref{acclPlate_CdPlot_linearAnisotropic}, we observe that as time progresses, the computed $C_D$ values obtained from the GPE+BV solver approach those of the reference solver. To provide a quantitative estimate of the error, we listed $C_D$ obtained from GPE+BV and the reference INS simulation in table~\ref{thinPlateDragCoeffComparisonTable}. For the entire time range reported, our method provides an accurate drag coefficient.

\begin{table}[!ht]
	\centering
	\caption{Percentage deviation of drag coefficient computed using GPE+BV compared to that obtained from INS.}
	\begin{tabular}{|cccc|}
		\hline
		\hline
		$\boldsymbol{t}$ & $\boldsymbol{C_{D}}$ (\textbf{INS})  & $\boldsymbol{C_{D}}$ (\textbf{GPE+BV})        & \% \textbf{Deviation}  \\
		\hline
            2 & 2.2833 & 2.2787 & 0.199 \\
            3 & 2.0106 & 2.0341 & 1.167 \\
            4 & 1.8881 & 1.9089 & 1.102 \\
            5 & 1.8206 & 1.8327 & 0.665  \\
            6 & 1.7707 & 1.7822 & 0.648 \\
            7 & 1.7357 & 1.7443 & 0.497 \\
            8 & 1.7096 & 1.7168 & 0.422 \\
            9 & 1.6894 & 1.6955 & 0.358 \\
            10 & 1.6734 & 1.6786 & 0.314 \\
		\hline
		\hline
	\end{tabular}
	\label{thinPlateDragCoeffComparisonTable}
\end{table}

\begin{figure}[H]
    \begin{center}
        \subfigure[]
        {
			\includegraphics[trim = 0mm 0mm 0mm 0mm, clip, width=15cm]{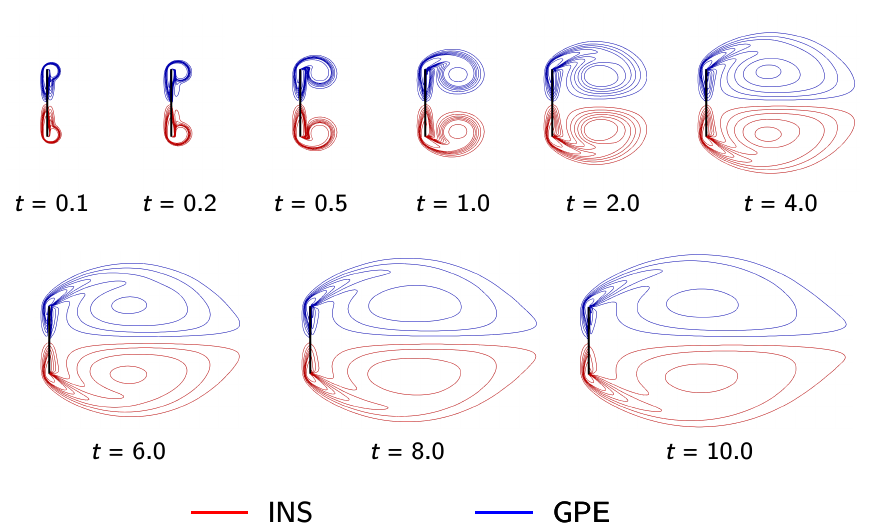}
        }
        \subfigure[]
        {
			\includegraphics[trim = 0mm 0mm 0mm 0mm, clip, width=15cm]{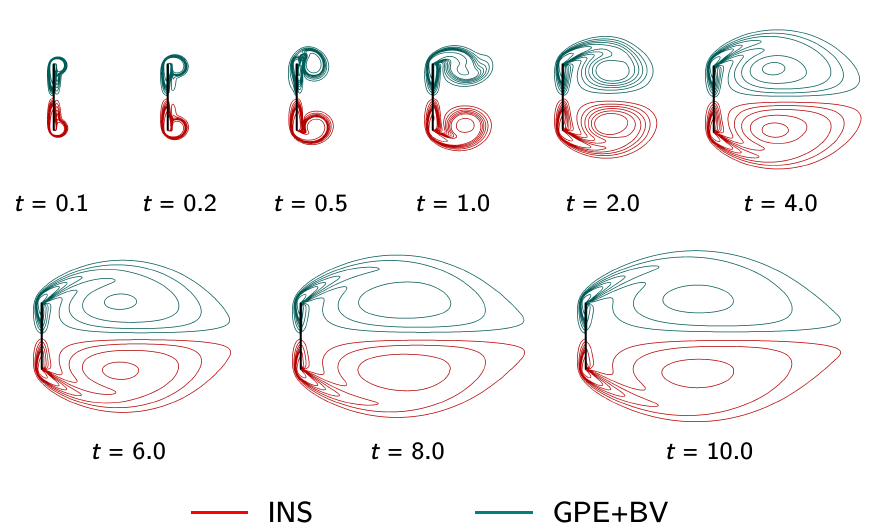}
        }
    \end{center}
	\caption{Comparison of vorticity contours from (a)~GPE and 
 (b)~GPE+BV with the INS solver, at various time instances for the flow over an impulsively started thin plate test case. The contours plotted are for $\omega_z = \pm1, \pm2, \pm3, \pm4, \pm5, \pm6, \pm10$.}
	\label{accl_vorticityContours}
\end{figure}

To qualitatively analyse the unsteady flow field, the evolution of vorticity contours is illustrated in figure~\ref{accl_vorticityContours}. These plots enable a robust one-to-one comparison between results obtained from GPE and GPE+BV with those from the INS solver. We observe that two counter-rotating vortices appear in the wake region behind the plate. Over time, these vortices increase in size as they are constantly fed by the shear layer appearing at the tip of the plate~\cite{koumoutsakos1996}. The contours from the GPE+BV solver exhibit a slight mismatch during the initial transients (before $t = 2$). However, an excellent agreement between these results can be seen after $t=2$ when sufficient acoustic damping has been achieved. A similar observation is made by Sun et al.~\cite{sun2023} for the viscous flow past an inclined elliptic cylinder problem where the lift and drag coefficient plots exhibit high fluctuations within non-dimensional time of $10$, even with the inclusion of an (isotropic) acoustic damper term in their smoothed-particle hydrodynamics solver.

\begin{figure}[!ht]
	\centering
	\subfigure[]
	{
		\includegraphics[trim = 5mm 0mm 11mm 5mm, clip, width=7.5cm]{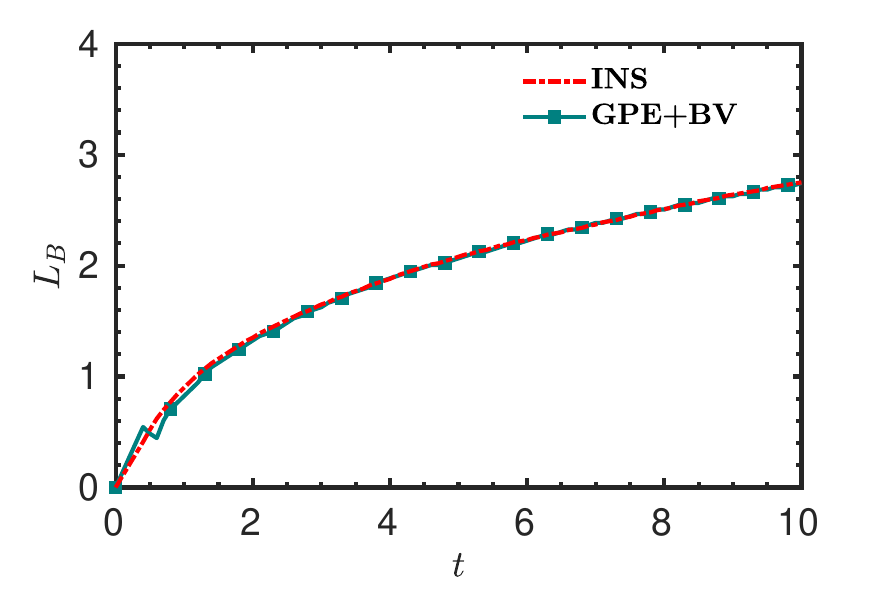}
		\label{acclPlate_bubbleLength_GPE_Re1000}
	}
	\subfigure[]
	{
		\includegraphics[trim = 5mm 0mm 11mm 5mm, clip, width=7.5cm]{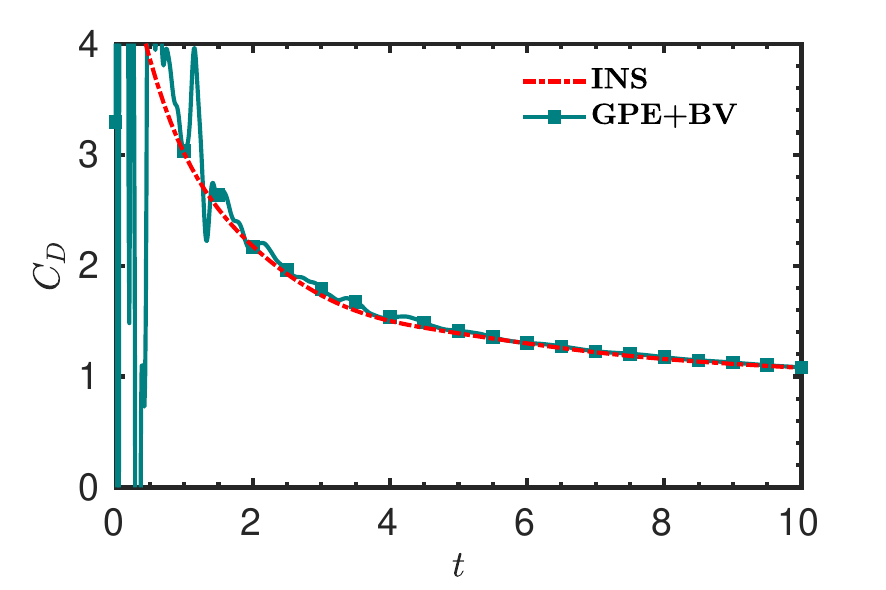}
		\label{acclPlate_drag_GPE_Re1000}
	}
	\caption{Evolution of (a) bubble length and (b) drag coefficient for the flow over an impulsively started thin plate test case, with $Re=1000$. }
	\label{acclPlate_gpe_Re1000}
\end{figure}

To provide additional confirmation for the effectiveness of the proposed approach, we investigate its performance at a higher $Re$. We simulated the impulsively started plate at $Re=1000$ on the same mesh and with the same time-step using GPE and GPE+BV solver with non-homogeneous anisotropic bulk viscosity; the evolution of $L_B$ and $C_D$ are plotted in figure~\ref{acclPlate_gpe_Re1000}. The accuracy of our method at this high $Re$ is also evident from the figure. This proves the robustness of the proposed approach. Similar to $Re=126$, the isotropic variants of bulk viscosity were ineffective.

It is to be noted that, compared to the damping mechanism using pressure diffusion, the bulk viscosity approach is better suited for high Re flows. This is because the magnitude of the pressure diffusion, which is decided by its coefficient, $1/Re Pr$, reduces when $Re$ increases, whereas $\boldsymbol{\mathcal{B}}$ is unaltered with higher $Re$.

For the effective damping of acoustic waves, the time scale associated with bulk viscosity should be smaller than the acoustic time scale. In contrast, for the above examples, we chose both these time scales of the same order so that the same $\Delta t$ can be used for both GPE and GPE+BV solvers. Despite this fact, the non-homogeneous anisotropic GPE+BV variant suppresses acoustic waves effectively, except for $t<2$. Although this can be viewed as a drawback of the present scheme, we highlight that our approach outperforms the isotropic bulk viscosity that is widely used in weakly compressible solvers.

The test cases presented in this paper are confined to two-dimensional problems. The extension of the present formulation to three-dimensional Cartesian grids is straightforward, and we anticipate similar advantages demonstrated for 2D cases. However, similar to AbdulGafoor et al.~\cite{abdulgafoor2024}, the investigation of the influence of the anisotropic bulk viscosity proposed in this paper on unstructured and moving/deforming grids requires a separate study.

\section{Conclusions}
  \label{conclusion_section}
Weakly compressible approaches to simulate incompressible flows have gained a lot of popularity in recent years because of their potential to achieve scalability in HPC implementations. A major limitation of such methods is the presence of artificial acoustic waves that act as the source of large mass conservation errors and highly oscillatory pressure fields. Although pressure diffusion and bulk viscosity are employed in many works to address this issue, their relative effectiveness remains unclear. In this paper, we demonstrated using various numerical examples that the bulk viscosity is more effective in annihilating acoustic waves than the pressure diffusion. We presented a comprehensive study of bulk viscosity on various aspects of weakly compressible methods, including damping of pressure oscillations, order of convergence and mass conservation error. A key contribution of this work is the formulation of a non-homogeneous anisotropic form of the bulk viscosity. We proved that the chosen form exhibited superior performance in annihilating the acoustic waves, even in the initial transients of unsteady flows. Such an example is highly challenging for weakly compressible methods, and we underline that the proposed bulk viscosity was significantly better in predicting the forces acting on a solid body. We carried out our investigations using a specific weakly compressible method called the general pressure equation. However, our findings are not restricted to this method alone, as the results we obtained from the entropically damped artificial compressibility based solver (not presented in this paper) also lead to similar conclusions. Moreover, bulk viscosity has been commonly used in the lattice Boltzmann methods and smoothed particle hydrodynamics based solvers. We envisage that these widely used weakly compressible methods will also benefit from the non-homogeneous anisotropic form of the bulk viscosity proposed in this paper.

\section*{CRediT authorship contribution statement}
\textbf{Dheeraj Raghunathan:} Data curation, Investigation, Methodology, Software, Formal analysis, Validation, Visualization, Writing – original draft. \textbf{Y. Sudhakar:} Conceptualization, Methodology,  Resources, Supervision, Writing – review \& editing.

\section*{Declaration of competing interest}
The authors declare that they have no known competing ﬁnancial interests or personal relationships that could have appeared to inﬂuence the work reported in this paper.

\section*{Data availability}
The data supporting this study’s findings are available from the corresponding author upon reasonable request.

\section*{Acknowledgments}
We acknowledge the financial support provided by IIT Goa in the form of a startup grant. Simulations are run on computational resources setup from the DST-SERB Ramanujan fellowship (SB/S2/RJN-037/2018). The first author would like to acknowledge the contribution endowed by Mr. Smruti Ranjan Jena for simulating the flow over an impulsively started thin plate using OpenFOAM.

\bibliographystyle{ieeetr}
\biboptions{sort&compress}
\bibliography{refs}

\begin{thebibliography}{10}

\bibitem{patankar}
S.~Patankar, {\em Numerical heat transfer and fluid flow}.
\newblock Taylor \& Francis, 2018.

\bibitem{Brown1995}
D.~L. Brown and M.~L. Minion, ``{Performance of under-resolved two-dimensional
  incompressible flow simulations},'' {\em Journal of Computational Physics},
  vol.~122, no.~1, pp.~165--183, 1995.

\bibitem{perot_1993}
J.~Perot, ``An analysis of the fractional step method,'' {\em Journal of
  Computational Physics}, vol.~108, no.~1, pp.~51--58, 1993.

\bibitem{Borok2007}
S.~Borok, S.~Ansumali, and I.~V. Karlin, ``{Kinetically reduced local
  Navier-Stokes equations for simulation of incompressible viscous flows},''
  {\em Physical Review E - Statistical, Nonlinear, and Soft Matter Physics},
  vol.~76, no.~6, pp.~1--9, 2007.

\bibitem{Chorin1967}
A.~J. Chorin, ``A numerical method for solving incompressible viscous flow
  problems,'' {\em Journal of Computational Physics}, vol.~2, no.~1,
  pp.~12--26, 1967.

\bibitem{turkel1987}
E.~Turkel, ``Preconditioned methods for solving the incompressible and low
  speed compressible equations,'' {\em Journal of Computational Physics},
  vol.~72, no.~2, pp.~277--298, 1987.

\bibitem{peyret1983}
R.~Peyret and T.~D. Taylor, {\em Computational methods for fluid flow}.
\newblock Springer Berlin Heidelberg, 1983.

\bibitem{merkle1987}
C.~Merkle, ``Time-accurate unsteady incompressible flow algorithms based on
  artificial compressibility,'' {\em 8th Computational Fluid Dynamics
  Conference}, 1987.

\bibitem{soh1987}
W.~Soh, ``Time-marching solution of incompressible {Navier}-{Stokes} equations
  for internal flow,'' {\em Journal of Computational Physics}, vol.~70, no.~1,
  pp.~232--252, 1987.

\bibitem{rogers1991upwind}
S.~E. Rogers and D.~Kwak, ``An upwind differencing scheme for the
  incompressible {N}avier--{S}tokes equations,'' {\em Applied Numerical
  Mathematics}, vol.~8, no.~1, pp.~43--64, 1991.

\bibitem{Kiris2002}
C.~Kiris and D.~Kwak, ``{Aspects of unsteady incompressible flow
  simulations},'' {\em Computers \& Fluids}, vol.~31, no.~4-7, pp.~627--638,
  2002.

\bibitem{nourgaliev2004}
R.~Nourgaliev, T.~Dinh, and T.~Theofanous, ``A pseudocompressibility method for
  the numerical simulation of incompressible multifluid flows,'' {\em
  International Journal of Multiphase Flow}, vol.~30, no.~7-8, pp.~901--937,
  2004.

\bibitem{Kim1999}
W.~W. Kim and S.~Menon, ``{An unsteady incompressible Navier-Stokes solver for
  large eddy simulation of turbulent flows},'' {\em International Journal for
  Numerical Methods in Fluids}, vol.~31, no.~6, pp.~983--1017, 1999.

\bibitem{ranjan2020}
R.~Ranjan, M.~K. Venkataswamy, and S.~Menon, ``Dynamic one-equation-based
  subgrid model for large-eddy simulation of stratified turbulent flows,'' {\em
  Physical Review Fluids}, vol.~5, no.~6, p.~064601, 2020.

\bibitem{Clausen2013}
J.~R. Clausen, ``{Entropically damped form of artificial compressibility for
  explicit simulation of incompressible flow},'' {\em Physical Review E},
  vol.~87, no.~1, pp.~1--12, 2013.

\bibitem{Toutant2017}
A.~Toutant, ``{General and exact pressure evolution equation},'' {\em Physics
  Letters, Section A: General, Atomic and Solid State Physics}, vol.~381,
  no.~44, pp.~3739--3742, 2017.

\bibitem{Toutant2018}
A.~Toutant, ``{Numerical simulations of unsteady viscous incompressible flows
  using general pressure equation},'' {\em Journal of Computational Physics},
  vol.~374, pp.~822--842, 2018.

\bibitem{kajzer2018}
A.~Kajzer and J.~Pozorski, ``Application of the {Entropically} {Damped}
  {Artificial} {Compressibility} model to direct numerical simulation of
  turbulent channel flow,'' {\em Computers \& Mathematics with Applications},
  vol.~76, no.~5, pp.~997--1013, 2018.

\bibitem{trojak2022}
W.~Trojak, N.~Vadlamani, J.~Tyacke, F.~Witherden, and A.~Jameson, ``Artificial
  compressibility approaches in flux reconstruction for incompressible viscous
  flow simulations,'' {\em Computers \& Fluids}, vol.~247, p.~105634, 2022.

\bibitem{dupuy2020}
D.~Dupuy, A.~Toutant, and F.~Bataille, ``{Analysis of artificial pressure
  equations in numerical simulations of a turbulent channel flow},'' {\em
  Journal of Computational Physics}, vol.~411, p.~109407, 2020.

\bibitem{Shi2020}
X.~Shi and C.~A. Lin, ``Simulations of wall bounded turbulent flows using
  general pressure equation,'' {\em Flow, Turbulence and Combustion}, vol.~105,
  no.~1, pp.~67--82, 2020.

\bibitem{kajzer2020}
A.~Kajzer and J.~Pozorski, ``A weakly compressible, diffuse-interface model for
  two-phase flows,'' {\em Flow, Turbulence and Combustion}, vol.~105, no.~2,
  pp.~299--333, 2020.

\bibitem{huang2020arxiv}
J.-J. Huang, ``Numerical simulation of two-phase incompressible viscous flows
  using general pressure equation,'' 2020.
\newblock arXiv:2011.00814 [physics].

\bibitem{bodhanwalla2023}
H.~Bodhanwalla, D.~Raghunathan, and Y.~Sudhakar, ``A general pressure equation
  based method for incompressible two-phase flows,'' {\em International Journal
  for Numerical Methods in Fluids}, vol.~96, no.~10, pp.~1653--1679, 2024.

\bibitem{sharma2023}
M.~Sharma, K.~Srikanth, T.~Jayachandran, and A.~Sameen, ``{DNS} of
  buoyancy-driven flows using {EDAC} formulation solved by high-order method,''
  {\em Computers \& Fluids}, vol.~265, p.~105997, 2023.

\bibitem{bolduc2023}
M.-P. Bolduc, R.~Ghoreishi, and B.~C. Vermeire, ``A high-order
  entropically-damped artificial compressibility approach on moving and
  deforming domains,'' {\em Computers \& Fluids}, vol.~257, p.~105839, 2023.

\bibitem{Ansumali2005}
S.~Ansumali, I.~V. Karlin, and H.~C. {\"{O}}ttinger, ``{Thermodynamic theory of
  incompressible hydrodynamics},'' {\em Physical Review Letters}, vol.~94,
  no.~8, pp.~1--4, 2005.

\bibitem{lu2023}
J.~Lu and N.~A. Adams, ``General mechanisms for stabilizing weakly compressible
  models,'' {\em Physical Review {E}}, vol.~107, p.~055306, 2023.

\bibitem{ramshaw1990}
J.~Ramshaw and V.~Mousseau, ``Accelerated artificial compressibility method for
  steady-state incompressible flow calculations,'' {\em Computers \& Fluids},
  vol.~18, no.~4, pp.~361--367, 1990.

\bibitem{mchugh1995}
P.~McHugh and J.~Ramshaw, ``Damped artificial compressibility iteration scheme
  for implicit calculations of unsteady incompressible flow,'' {\em
  International {J}ournal for {N}umerical {M}ethods in {F}luids}, vol.~21,
  no.~2, pp.~141--153, 1995.

\bibitem{ramshaw1991lowMach}
J.~Ramshaw and V.~Mousseau, ``Damped artificial compressibility method for
  steady-state low-speed flow calculations,'' {\em Computers \& {F}luids},
  vol.~20, no.~2, pp.~177--186, 1991.

\bibitem{mazaheri2003}
K.~Mazaheri and P.~L. Roe, ``Bulk viscosity damping for accelerating
  convergence of low {M}ach number {E}uler solvers,'' {\em International
  {J}ournal for {N}umerical {M}ethods in {F}luids}, vol.~41, no.~6,
  pp.~633--652, 2003.

\bibitem{yasuda2023}
T.~Yasuda, I.~Tanno, T.~Hashimoto, K.~Morinishi, and N.~Satofuka, ``Artificial
  compressibility method using bulk viscosity term for an unsteady
  incompressible flow simulation,'' {\em Computers \& Fluids}, vol.~258,
  p.~105885, 2023.

\bibitem{abdulgafoor2024}
C.~AbdulGafoor, A.~Rajananda, A.~Shankar, and N.~R. Vadlamani, ``Entropy
  damping and {Bulk} {Viscosity} based artificial compressibility methods on
  dynamically distorting grids,'' {\em Computers \& Fluids}, vol.~279,
  p.~106328, 2024.

\bibitem{dellar2001}
P.~J. Dellar, ``Bulk and shear viscosities in lattice {Boltzmann} equations,''
  {\em Physical Review E}, vol.~64, no.~3, p.~031203, 2001.

\bibitem{asinari2010}
P.~Asinari and I.~V. Karlin, ``Quasiequilibrium lattice {B}oltzmann models with
  tunable bulk viscosity for enhancing stability,'' {\em Physical Review E},
  vol.~81, p.~016702, 2010.

\bibitem{avalos2020}
J.~B. Avalos, M.~Antuono, A.~Colagrossi, and A.~Souto-Iglesias,
  ``Shear-viscosity-independent bulk-viscosity term in smoothed particle
  hydrodynamics,'' {\em Physical Review E}, vol.~101, no.~1, p.~013302, 2020.

\bibitem{sun2023}
P.~Sun, C.~Pilloton, M.~Antuono, and A.~Colagrossi, ``Inclusion of an acoustic
  damper term in weakly-compressible {SPH} models,'' {\em Journal of
  Computational Physics}, vol.~483, p.~112056, 2023.

\bibitem{ramshaw1986}
J.~Ramshaw, P.~O'Rourke, and A.~Amsden, ``Acoustic damping for explicit
  calculations of fluid flow at low {Mach} number,'' Tech. Rep. LA-10641-MS,
  6100813, University of North Texas Libraries, UNT Digital Library, 1986.

\bibitem{pan2022}
D.~Pan, ``A high-order finite volume method solving viscous incompressible
  flows using general pressure equation,'' {\em Numerical Heat Transfer, Part
  B: Fundamentals}, vol.~82, no.~5, pp.~146--163, 2022.

\bibitem{Parameswaran2019}
S.~Parameswaran and J.~C. Mandal, ``{A novel Roe solver for incompressible
  two-phase flow problems},'' {\em Journal of Computational Physics}, vol.~390,
  pp.~405--424, 2019.

\bibitem{yang2021}
K.~Yang and T.~Aoki, ``Weakly compressible {Navier}-{Stokes} solver based on
  evolving pressure projection method for two-phase flow simulations,'' {\em
  Journal of Computational Physics}, vol.~431, p.~110113, 2021.

\bibitem{kajzer2022}
A.~Kajzer and J.~Pozorski, ``A weakly compressible, diffuse interface model of
  two-phase flows: {Numerical} development and validation,'' {\em Computers \&
  Mathematics with Applications}, vol.~106, pp.~74--91, 2022.

\bibitem{Ghia1982}
U.~Ghia, K.~N. Ghia, and C.~T. Shin, ``{High-Re solutions for incompressible
  flow using the Navier-Stokes equations and a multigrid method},'' {\em
  Journal of Computational Physics}, vol.~48, no.~3, pp.~387--411, 1982.

\bibitem{he2002}
X.~He, G.~D. Doolen, and T.~Clark, ``Comparison of the lattice {Boltzmann}
  method and the artificial compressibility method for {Navier}–{Stokes}
  equations,'' {\em Journal of Computational Physics}, vol.~179, no.~2,
  pp.~439--451, 2002.

\bibitem{nagata2021}
K.~Nagata, N.~Ikegaya, and J.~Tanimoto, ``Consideration of artificial
  compressibility for explicit computational fluid dynamics simulation,'' {\em
  Journal of Computational Physics}, vol.~443, p.~110524, 2021.

\bibitem{bell1989}
J.~B. Bell, P.~Colella, and H.~M. Glaz, ``A second-order projection method for
  the incompressible {N}avier-{S}tokes equations,'' {\em Journal of
  Computational Physics}, vol.~85, no.~2, pp.~257--283, 1989.

\bibitem{minion1997}
M.~L. Minion and D.~L. Brown, ``Performance of under-resolved two-dimensional
  incompressible flow simulations, {II},'' {\em Journal of Computational
  Physics}, vol.~138, no.~2, pp.~734--765, 1997.

\bibitem{Shah2010}
A.~Shah, L.~Yuan, and A.~Khan, ``{Upwind compact finite difference scheme for
  time-accurate solution of the incompressible Navier-Stokes equations},'' {\em
  Applied Mathematics and Computation}, vol.~215, no.~9, pp.~3201--3213, 2010.

\bibitem{hashimoto2018}
T.~Hashimoto, T.~Yasuda, I.~Tanno, Y.~Tanaka, K.~Morinishi, and N.~Satofuka,
  ``Multi-{GPU} parallel computation of unsteady incompressible flows using
  kinetically reduced local {Navier}–{Stokes} equations,'' {\em Computers \&
  Fluids}, vol.~167, pp.~215--220, 2018.

\bibitem{de_mulder1998}
T.~De~Mulder, ``The role of bulk viscosity in stabilized finite element
  formulations for incompressible flow: {A} review,'' {\em Computer Methods in
  Applied Mechanics and Engineering}, vol.~163, no.~1-4, pp.~1--10, 1998.

\bibitem{rasthofer2018}
U.~Rasthofer and V.~Gravemeier, ``Recent developments in variational multiscale
  methods for large-eddy simulation of turbulent flow,'' {\em Archives of
  Computational Methods in Engineering}, vol.~25, no.~3, pp.~647--690, 2018.

\bibitem{olshanskii2009}
M.~Olshanskii, G.~Lube, T.~Heister, and J.~Löwe, ``Grad–div stabilization
  and subgrid pressure models for the incompressible {Navier}–{Stokes}
  equations,'' {\em Computer Methods in Applied Mechanics and Engineering},
  vol.~198, no.~49-52, pp.~3975--3988, 2009.

\bibitem{taylor1937}
G.~I. Taylor and A.~E. Green, ``Mechanism of the production of small eddies
  from large ones,'' {\em Proceedings of the Royal Society of London. Series
  A-Mathematical and Physical Sciences}, vol.~158, no.~895, pp.~499--521, 1937.

\bibitem{sharma2004a}
A.~Sharma and V.~Eswaran, ``Heat and fluid flow across a square cylinder in the
  two-dimensional laminar flow regime,'' {\em Numerical Heat Transfer, Part A:
  Applications}, vol.~45, no.~3, pp.~247--269, 2004.

\bibitem{sohankar1998}
A.~Sohankar, C.~Norberg, and L.~Davidson, ``Low-{Reynolds}-number flow around a
  square cylinder at incidence: study of blockage, onset of vortex shedding and
  outlet boundary condition,'' {\em International Journal for Numerical Methods
  in Fluids}, vol.~26, no.~1, pp.~39--56, 1998.

\bibitem{sen2011}
S.~Sen, S.~Mittal, and G.~Biswas, ``Flow past a square cylinder at low
  {Reynolds} numbers,'' {\em International Journal for Numerical Methods in
  Fluids}, vol.~67, no.~9, pp.~1160--1174, 2011.

\bibitem{zhao2013}
M.~Zhao, L.~Cheng, and T.~Zhou, ``Numerical simulation of vortex-induced
  vibration of a square cylinder at a low {Reynolds} number,'' {\em Physics of
  Fluids}, vol.~25, no.~2, p.~023603, 2013.

\bibitem{saha2015}
A.~K. Saha and A.~Shrivastava, ``Suppression of vortex shedding around a square
  cylinder using blowing,'' {\em Sadhana}, vol.~40, no.~3, pp.~769--785, 2015.

\bibitem{taneda1971}
S.~Taneda and H.~Honji, ``Unsteady flow past a flat plate normal to the
  direction of motion,'' {\em Journal of the Physical Society of Japan},
  vol.~30, no.~1, pp.~262--272, 1971.

\bibitem{koumoutsakos1996}
P.~Koumoutsakos and D.~Shiels, ``Simulations of the viscous flow normal to an
  impulsively started and uniformly accelerated flat plate,'' {\em Journal of
  Fluid Mechanics}, vol.~328, pp.~177--227, 1996.

\end{thebibliography}

\appendix

\section{Significance of the additional bulk viscosity term}
\label{appendix_secondTerm}
\begin{figure}[!ht]
	\begin{center}
		\includegraphics[trim = 0mm 0mm 0mm 0mm, clip, width=15.0cm]{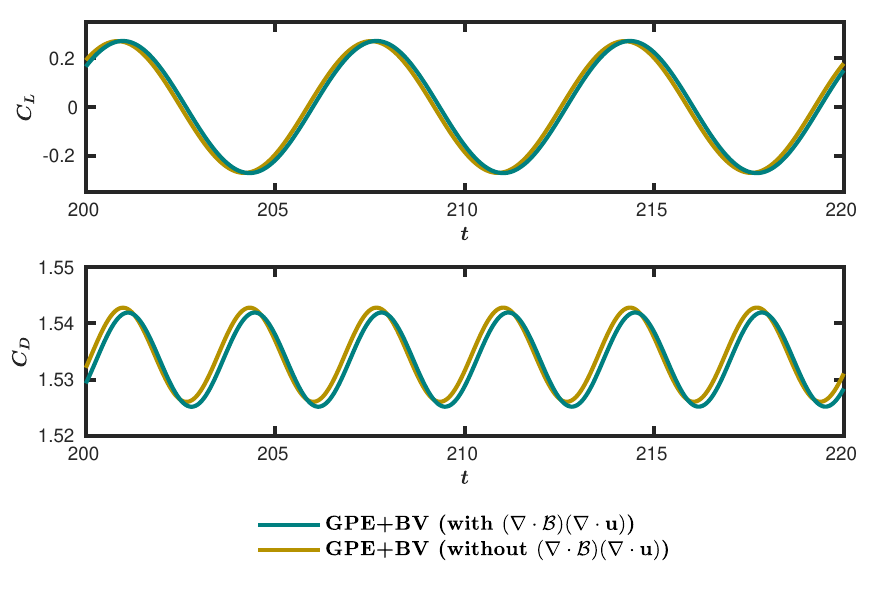}
	\end{center} 
	\caption{Effect of second bulk viscosity term for the flow over a square cylinder test case.}
	\label{cylinder_addTerm}
\end{figure}

\begin{figure}[!ht]
	\begin{center}
		\includegraphics[trim = 0mm 0mm 0mm 0mm, clip, width=10.0cm]{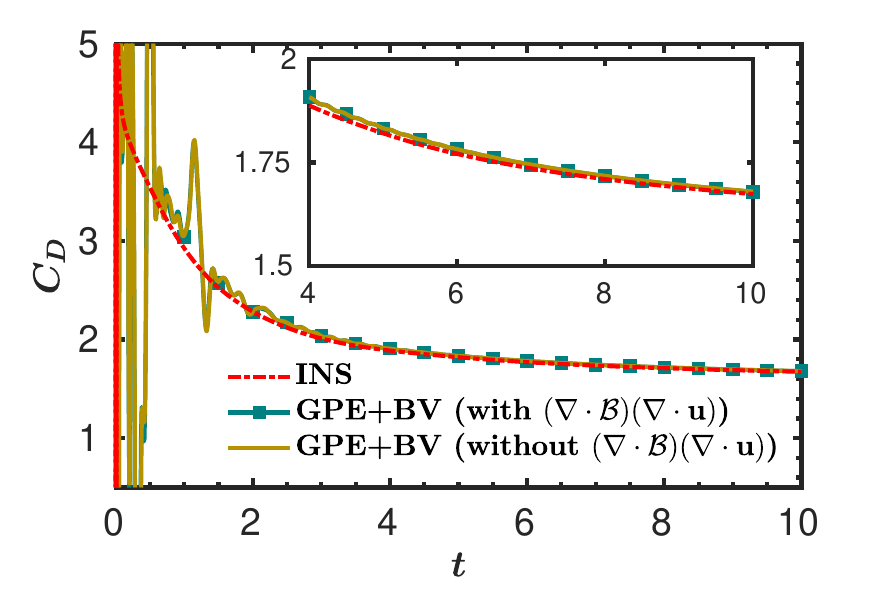}
	\end{center} 
	\caption{Effect of second bulk viscosity term for the flow over an impulsively started thin plate test case.}
	\label{thinPlate_drag_addTerm}
\end{figure}

Due to its non-homogeneous nature, the artificial bulk viscosity term expands to $\nabla \cdot \left( \boldsymbol{\mathcal{B}} \nabla \cdot \mathbf{u} \right) =  \boldsymbol{\mathcal{B}} \nabla \left( \nabla \cdot \mathbf{u} \right) +    (\nabla \cdot \boldsymbol{\mathcal{B}}) \left(  \nabla \cdot \mathbf{u} \right)$. Here, the second term $(\nabla \cdot \boldsymbol{\mathcal{B}}) \left(  \nabla \cdot \mathbf{u} \right)$ is zero for a uniform mesh; on a non-uniform mesh, its magnitude depends on the stretching ratio, which, for all practical computations, is maintained close to 1. This implies $\nabla \cdot \boldsymbol{\mathcal{B}}\ll 1$; hence, this term makes only a non-significant contribution. In this section, we examine this presumption using the two test cases for which we use stretched grids. Figure~\ref{cylinder_addTerm} shows the $C_L$ and $C_D$ plots during the periodic phase obtained from the simulations of flow over a square cylinder, with and without the additional term. Similarly, in figure~\ref{thinPlate_drag_addTerm}, we present the evolution of $C_D$ for the case of flow over an impulsively started thin plate. In both cases, we observe only negligible contribution from $(\nabla \cdot \boldsymbol{\mathcal{B}}) \left(  \nabla \cdot \mathbf{u} \right)$. This is also evident from table~\ref{addTermComparisonTable}.

\begin{table}[h!]
    \centering
    \caption{ Comparison of simulation data with and without $(\nabla \cdot \boldsymbol{\mathcal{B}}) \left(  \nabla \cdot \mathbf{u} \right)$ for the flow over an impulsively started thin plate test case.  } 
    \begin{tabular}{|c|c|c|c|}
            \hline
            \hline
            \multicolumn{4}{|c|}{\textbf{Square Cylinder}} \\
            \hline
           \textbf{Parameter} & \textbf{with} $(\nabla \cdot \boldsymbol{\mathcal{B}}) \left(  \nabla \cdot \mathbf{u} \right)$ & 
           \textbf{without} $(\nabla \cdot \boldsymbol{\mathcal{B}}) \left(  \nabla \cdot \mathbf{u} \right)$ & \% \textbf{Deviation}  \\
            \hline
           $C_{D_{avg}}$ & 1.5336 & 1.5344 & 0.052 \\
           \hline
           $ C_{L_{rms}} $ & 0.1924 & 0.1917 & 0.36 \\
           \hline
           \multicolumn{4}{|c|}{\textbf{Impulsively started plate}} \\
           \hline
           $C_D$ at $t=10$ &  1.678682 & 1.679478 & 0.047 \\
        \hline
        \hline
    \end{tabular}
    \label{addTermComparisonTable}
\end{table}
 
\section{Effect of pressure diffusion term on damping the acoustic waves}
\label{appendix_PrStudy}

\begin{figure}[!ht]
	\begin{center}
		\includegraphics[trim = 0mm 0mm 0mm 0mm, clip, width=15.0cm]{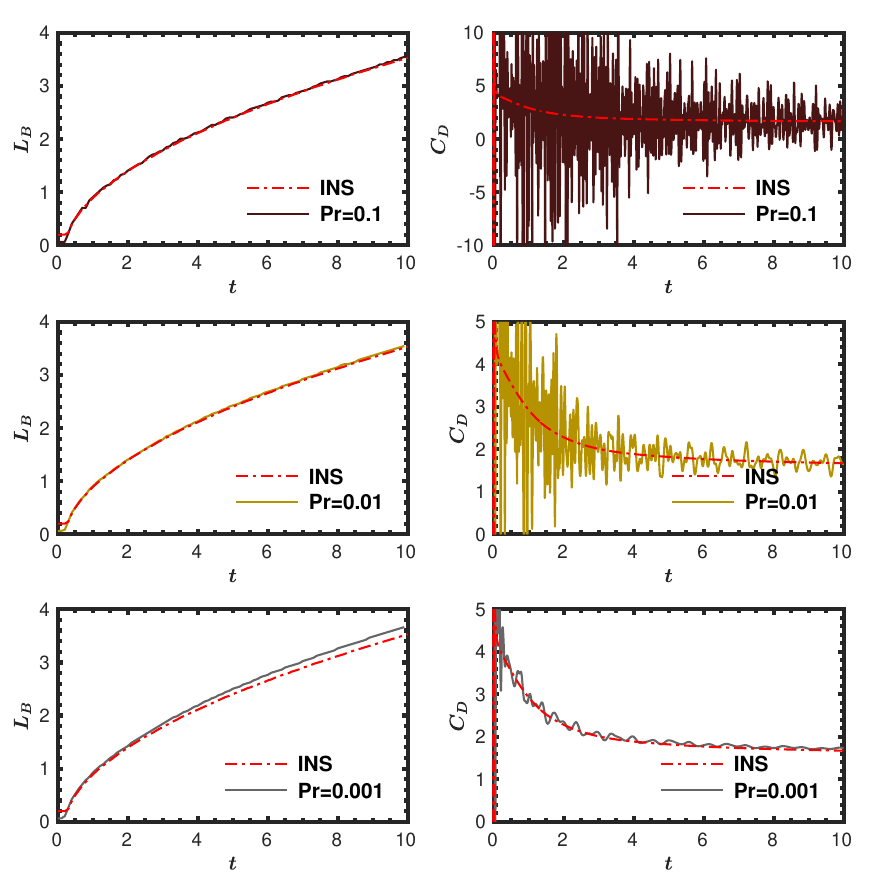}
	\end{center} 
	\caption{Effect of Prandtl number on the evolution of bubble length ($L_B$) and drag coefficient ($C_D$) for the flow over an impulsively started thin plate test case.}
	\label{thinPlate_bubbleLength_PrStudy}
\end{figure} 
We conducted a detailed parametric study on $Pr$ for the flow over impulsively started thin plate case. The values used are $Pr$=0.1, 0.01 and 0.001, and the plots of $L_B$ and $C_D$ with respect to time are presented in figure~\ref{thinPlate_bubbleLength_PrStudy}. From the plots, we conclude that the acoustic waves reduce as $Pr$ reduces. The most effective damping is observed for the lowest value of $Pr$ ($Pr=0.001$). However, the corresponding $L_B$ curve deviates from the reference values. The maximum error for $L_B$ at $t=10$ is estimated to be 4.4\%. Moreover, considering stability issues, reducing $Pr$ results in a compulsory need to reduce the time step. For the simulations with $Pr=0.1$ and $Pr=0.01$, the time-step required is $10^{-5}$ and $10^{-6}$, respectively, whereas, for $Pr=0.001$, we need to reduce it further to $10^{-7}$. Evidently, the rate of damping with $Pr=0.001$ is still not sufficient, and further reducing $Pr$ results in both increased computational cost and reduced accuracy. To reiterate, any attempt to damp the acoustic waves by adjusting the numerical parameters in GPE negates the advantage of GPE over other conventional schemes in terms of computational efficiency. 

\end{document}